\documentclass[aps,prc,preprint,tightenlines,superscriptaddress,showpacs,%
amssymb,byrevtex,nofootinbib]{revtex4}
\usepackage[dvips]{graphicx,color}
\usepackage{dcolumn}
\usepackage{epsfig}
\usepackage{bm}

\begin{document}
\title{Photo-production of Bound States with Hidden Charms }
\author{ Jia-Jun Wu}
\affiliation{Physics Division, Argonne National Laboratory, Argonne, Illinois 60439, USA}
\author{T.-S. H. Lee}
\affiliation{Physics Division, Argonne National Laboratory, Argonne, Illinois 60439, USA}
\begin{abstract}

The photo-production of $J/\Psi$-$^3He$ bound state ($[^3He]_{J/\Psi}$) on
a $^4He$ target has been investigated
 using the impulse approximation. The calculations have been performed  using several  $\gamma+N \rightarrow J/\Psi +N$ models based on the Pomeron-exchange and accounting for the pion-exchange mechanism at low energies. The $J/\Psi$ wavefunctions in $[^3He]_{J/\Psi}$ are generated from various $J/\Psi$-nucleus
potentials which are constructed by  either using a procedure
based on the Pomeron-quark coupling mechanism or folding a
$J/\Psi$-N potential ($v_{J/\Psi,N}$) into the nuclear densities. We consider $v_{J/\Psi,N}$ derived from the effective field theory approach, Lattice QCD, and Pomeron-quark coupling mechanism. The upper bound of the predicted total cross sections is about
$0.1 - 0.3$  pico-barn. We also consider the possibility of
photo-production of a six quark-$J/\Psi$ bound state ($[q^6]_{J/\Psi})$ on the $^3He$ target.
 The Compound Bag Model of $NN$ scattering and the quark cluster model of nuclei are used to estimate  the $[q^6]$-N wavefunction in $^3He$ by imposing the condition that the calculated $^3He$ charge form factor must be consistent with
what is predicted by the conventional nuclear model. The upper bound of the predicted total cross sections of $\gamma + ^3He \rightarrow [q^6]_{J/\Psi} +N$ is about 2 - 4 pico-barn, depending on the model of $\gamma+N \rightarrow J/\Psi +N$ used in the calculations. Our results call for the need of precise measurements of $\gamma+p \rightarrow J/\Psi +p$ and also the $\gamma+^2H\rightarrow J/\Psi +n + p $ reactions near the threshold.
\end{abstract}

\maketitle

\section{introduction}
The role of the gluon field in determining the interactions between nucleons
and quark-antiquark ($Q\bar{Q}$) systems, which do not share the same $up$ and $down$ quarks with the nucleon, is one of the
interesting subjects in understanding
Quantum Chromodynamics(QCD). An important  step toward this direction was taken by Peskin\cite{pesk79} who applied the methodology of the operator product
expansion to evaluate the strength of the color field emitted by
heavy $Q\bar{Q}$ systems. His results suggested\cite{bp79} that the van der Waals
force induced by the color field of $J/\Psi$ on nucleons can generate an attractive short-range $J/\Psi$-$N$
interaction. By using the effective field theory method, Luke, Manohar, and Savage\cite{luke92} used the results
from Peskin to predict the $J/\Psi$-nucleon forward scattering amplitude which was used to get an estimation that $J/\Psi$ can have a few MeV/nucleon attraction in nuclear matter. Brodsky and Miller\cite{brodsky-1} further
investigated the $J/\Psi$-N forward scattering amplitude of Ref.\cite{luke92}
to derive a $J/\Psi$-N potential ($v_{J/\Psi,N}$) which
gives an $J/\Psi$-N scattering length of -$0.24$ fm.
The result of Peskin was also used by Kaidalov and Volkovitsky\cite{russia},
 who differed from Ref.\cite{brodsky-1} in evaluating the gluon content in
the nucleon, to give a much smaller scattering length of -$0.05$ fm.
In a Lattice QCD calculation, Kawanai and Saski\cite{lqcd} obtained an
attractive $J/\Psi$-N potential $v_{J/\Psi,N} =$  - $  \alpha  e^{-\mu r}/{r}$
with $\alpha =  0.1$ and $\mu = 0.6$ GeV, which gives a scattering
length - $ 0.09$ fm. In Ref.\cite{brodsky90},
Brodsky, Schmidt
and de Teramond proposed an approach to calculate the  potential between a
$c\bar{c}$ meson and a nucleus by using the Pomeon-exchange model of Dannachie and Lanshoff\cite{donn84}. The $J/\Psi$-N potential obtained in this
approach is $v_{J/\Psi,N} =$ - $ \alpha e^{-\mu r}/{r}$ with $\alpha = 0.6$
and $\mu = 0.6$ GeV which gives a rather large scattering length - $ 8.83$ fm.

Our first objective in this paper is to explore whether these $J/\Psi$-N potentials, with rather different attractive strengths, can form $J/\Psi$-nucleus bound states. Following the well developed method in nuclear reaction theory\cite{feshbach-1}, this is done by searching for bound states by solving the Schrodinger equation with a folding potential constructed by integrating the $J/\Psi$-N potential over the nuclear density. We will also consider
the approach of Ref.\cite{brodsky90} in predicting $J/\Psi$-nucleus bound states by the coherent sum of Pomeon-exchange between quarks
in $J/\Psi$ and all quarks in the nucleus. For each of
the predicted bound $[^3He]_{J/\Psi}$ systems,
 we then estimate the photo-production cross section
of the $\gamma+ ^4He\rightarrow [^3He]_{J/\Psi} +N$ reaction
to facilitate  future experimental investigations\cite{workshop}.

The second part of this work is motivated by the investigations by Brodsky and de Teramond\cite{brodsky-3} who found that the spin correlation of $pp$ elastic scattering near the $J/\Psi$ production threshold can be explained if one postulates the excitation of a hidden charm ($c$) state $|qqqqqqc\bar{c}>$.
 Based on the similar consideration on the role of multi-quark configurations, Brodsky, Chudakov, Hoyer, and Laget\cite{brodsky-2}
suggested in a study of $\gamma + ^2H \rightarrow  J/\Psi+n+p$
reaction that $J/\Psi$ can interact strongly with the
six-quark $[q^6]$ component
of the deuteron wavefunction because the octet 3-quark $[q^3]_8$ in the $[q^6]$ can directly interact with each quark in $J/\Psi$. These two works suggest the possibility that if $J/\Psi$ overlap with a $[q^6]$ cluster in nuclei, a bound $[q^6]_{J/\Psi}$ system could be formed. It is of course very difficult, if not impossible,  to estimate $[q^6]$-${J/\Psi}$ interaction. Instead, we will simply assume the existence of such states and use the previous works\cite{low,mulder,itep,fasano-1,fasano-2,bakker,vary,japan-he3} on quark clusters in nuclei to explore how the cross sections of $\gamma + ^3He \rightarrow [q^6]_{J/\Psi} +N$ depend on the parameters characterizing the $[q^6]$-${J/\Psi}$ interaction within a potential model.

\begin{figure}[h]
\centering
\epsfig{file=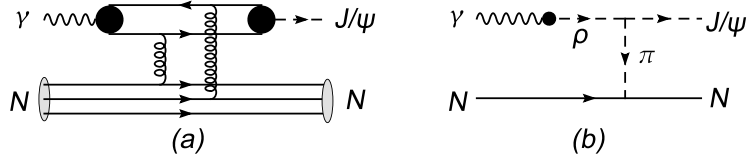, width=0.6\hsize}
\caption{Reaction mechanisms of
$\gamma N \rightarrow J/\Psi +N$ : (a) Pomeron-exchange,
(b) pion-exchange. }
\label{fig:pom-mec}
\end{figure}

Our first task is to construct a model of $\gamma + N \rightarrow J/\Psi +N$ reaction. At high energies, it is well recognized that this reaction can be described by the Pomeron-exchange model with an interpretation\cite{donn84,LN87,LM95,PL97} that Pomeron-exchange is due to the exchange of gluons within QCD.
This is illustrated in part (a) of Fig.\ref{fig:pom-mec}.
At low energies, one expects that  mechanisms other than Pomeron-exchange could also contribute as  can be seen in the exclusive $\phi$ photo-production
reaction on the nucleon\cite{PL97,titov-lee,ky12}.
However, very little investigation has been done for $J/\Psi$ photo-production in the near threshold region. As a first step, we will only consider the meson-exchange mechanism which can be calculated from using the partial decay width of $J/\Psi\to \pi \rho$ listed by Particle Data Group\cite{pdg} (PDG). With the vector meson dominance (VDM) assumption, this observed decay process indicates that $J/\Psi$ photo-production can also be due to the exchanges of a $\pi$ meson with the nucleon, as illustrated
in  part (b) of Fig.\ref{fig:pom-mec}.

We next consider the photo-nuclear reaction mechanism that a $J/\Psi$ is produced from a nucleon in a nucleus with mass number $A$ and then forms a bound state with the spectator $B$ system which can be a $(A-1)$ nuclear system
or a quark cluster  $[q^{3(A-1)}]$ in the target nucleus $A$. With this commonly used impulse approximation, the reaction cross sections can be calculated from the $\gamma + N \rightarrow J/\Psi +N$ amplitude, which will be generated from the Pomeron-exchange and pion-exchange mechanisms described above, and the initial nucleon and final $J/\Psi$ wavefunctions.
 The nucleon wavefunctions
 can be taken from the available nuclear models. The $J/\Psi$
  wavefunctions will be generated from various $J/\Psi$-$B$ potentials mentioned above. For simplicity, we only present the predictions of the cross sections of $\gamma+ [^4He] \rightarrow N +  [^3He]_{J/\Psi} $ and $\gamma+ [^3He] \rightarrow N +  [q^6]_{J/\Psi} $ reactions.

In section II, we present formula for calculating
the $\gamma + N \rightarrow J/\Psi +N$ amplitudes from the
Pomeron-exchange and pion-exchange mechanisms.
 The impulse approximation formula for calculating the cross sections of $\gamma + [A] \rightarrow [B]_{J/\Psi} +N$ are given in section III. Our results are presented in section IV.
In section V, we give a summary and discuss the necessary future work.

\section{Formula for $\gamma + N \rightarrow J/\Psi +N$ reaction}

In the center of mass system, the differential cross section of
$\gamma(\vec{q}) + N(-\vec{q}) \rightarrow J/\Psi(\vec{k}) + N(-\vec{k})$
with invariant mass $W$ can be written
\begin{eqnarray}
\frac{d\sigma}{d\Omega}&=&[\frac{(2\pi)^4
E_N(q)}{W}][\frac{kE_{J/\Psi}(k)E_N(k)}{W}]
\frac{1}{4}\sum_{\lambda_\gamma,\lambda_{J/\Psi}}\sum_{m_s,m'_s}
|<\vec{k}\lambda_{J/\Psi}m'_s|t(W)|\vec{q}\lambda_\gamma m_s>|^2 \,,
\end{eqnarray}
where $\lambda_{J/\Psi}$ and $\lambda_\gamma$ are the helicities of
the $J/\Psi$ and photon, respectively, $m_s$ is the z-component of the
nucleon spin,
and $E_a(p)=[m^2_a+\vec{p}^{\,\,2}]^{1/2}$ is the energy of a particle with
mas $m_a$. The reaction amplitude is written as
\begin{eqnarray}
<\vec{k}\lambda_{J/\Psi}m'_s|t(W)|\vec{q}\lambda_\gamma m_s>
& =&
\frac{1}{(2\pi)^3}\frac{1}{\sqrt{2E_{J/\Psi}(k)}}
\sqrt{\frac{m_N}{E_N(k)}}\sqrt{\frac{m_N}{E_N(q)}}\frac{1}{\sqrt{2q}}\nonumber \\
&&\times
[\bar{u}_{m'_s}(p')
\epsilon^*_\mu(k,\lambda_{J/\Psi})M^{\mu\nu}(p',p) \epsilon_\nu(q,\lambda_\gamma)
u_{m_s}({p})]\,, \label{eq:tot-amp}
\end{eqnarray}
where $u_{m_s}({p})$ is the nucleon spinor (with the normaliztion
$\bar{u}_{m_s}({p}){u}_{m'_s}({p}) = \delta_{m_s,m'_s}$) ,
$\epsilon_\mu(k,\lambda_{J/\Psi})$ and $\epsilon_\nu(q,\lambda_\gamma)$
are the polarization vectors of $J/\Psi$ and photon, respectively.
Here we also have introduced the four-momenta for the initial and final nucleons:
\begin{eqnarray}
p&=&(E_N(q), -\vec{q})\,\,\,\,\,;\,\,\, p'=(E_N(k),- \vec{k}) \,.\nonumber
\end{eqnarray}
In the following subsections, we give formula for calculating the
invariant amplitude $M^{\mu\nu}$
due to the Pomeron-exchange and meson-exchange mechanisms, as illustrated in
Fig.\ref{fig:pom-mec}.

\subsection{Pomeron-exchange amplitude}

Within the  Pomeron-exchange  model of
Donnachie and Landshoff \cite{donn84}, the vector meson photo-production at high energies is due to the mechanism that
 the incoming photon couples with a $q\bar{q}$ pair which interacts with the nucleon by the Pomeron exchange before
forming the outgoing vector meson. The quark-Pomeron vertex is obtained by the Pomeron-photon analogy\cite{donn84}, which treats the Pomeron as a $C=+1$ isoscalar photon, as
suggested by a study of non perturbative two-gluon exchanges \cite{LN87}.
Following the formula given explicitly in Ref.\cite{OL02},
we then have
\begin{equation}
\mathcal{M}^{\mu\nu}_\mathbb{P}(p',p) = G_\mathbb{P}^{}(s,t)
\mathcal{T}^{\mu\nu}_\mathbb{P}(p',p)
\label{eq:MP}
\end{equation}
with
\begin{equation}
\mathcal{T}^{\mu\nu}_\mathbb{P}(p',p) = i 12 \sqrt{4\pi\alpha_{\rm em}}
\frac{M_V^2 \beta_q \beta_{q'}}{f_V^{}} \frac{1}{M_V^2-t} \left(
\frac{2\mu_0^2}{2\mu_0^2 + M_V^2 - t} \right) F_1(t)
\{ k\!\!\!/ \, g^{\mu\nu} - k^\mu \gamma^\nu \}\, ,
\label{eq:pom}
\end{equation}
where $t=(p-p')^2$, $s=(q+p)^2=W^2$, $\alpha_{\rm em} = e^2/4\pi$, $\beta_q$ is the Pomeron-quark coupling
constant, $M_V^{}$ is the vector meson mass, and $F_1(t)$ is the isoscalar
electromagnetic form factor of the nucleon,
\begin{equation}
F_1(t) = \frac{4M_N^2 - 2.8 t}{(4M_N^2 - t)(1-t/0.71)^2}.
\end{equation}
Here $t$ is in unit of GeV$^2$, and $M_N$ is the proton mass.

The Regge propagator for the Pomeron in Eq. (\ref{eq:MP}) is
\begin{equation}
G^{}_\mathbb{P} = \left(\frac{s}{s_0^{}}\right)^{\alpha^{}_P(t)-1}
\exp\left\{ - \frac{i\pi}{2} \left[ \alpha_P^{}(t)-1 \right]
\right\} \,,
\label{eq:regge-g}
\end{equation}
where $ \alpha_P^{} (t) = \alpha_0 + \alpha'_P t$.
It is common\cite{OL02}  to
use $\alpha_0= 1.08$ and $\alpha'_P = 1/s_0^{} = 0.25$
GeV$^{-2}$.
In Eq.~(\ref{eq:pom}), $f_V$ is the vector meson decay constant:
$f_\rho = 5.33$, $f_\omega = 15.2$, $f_\phi = 13.4$, and $f_{J/\Psi} = 11.2$.
The other parameters in Eq.(\ref{eq:pom}) have been determined by
fitting\cite{OL02} the total cross section data of the
photo-production of $\rho$ and  $\omega$ :
$\beta_u=\beta_d = 2.07$ GeV$^{-1}$ and $\mu_0^2 = 1.1$ GeV$^2$.

With the parameters specified above, our task is to examine the extent
to which the total cross section of photo-production of $J/\Psi$ can be fitted by only
adjusting the Pomeron-charmed quark coupling constant $\beta_c$. This will be discussed in section IV.

\subsection{Pion-exchange amplitude}
We observe from  Particle Data\cite{pdg}
that the width of the $J/\Psi \to \pi^0\rho^0$ is significant,
\begin{eqnarray}
\Gamma_{J/\Psi \rightarrow \pi^0\rho^0} = 0.92\,\,\,MeV\,\,\,
\times (0.56\pm 0.07)\% \,.  \label{eq:jpsi-width}
\end{eqnarray}
With the vector meson dominance (VDM) assumption,
this experimental information
 allows us to calculate
the one-pion-exchange amplitude of $\gamma + N \rightarrow J/\Psi +N$,
as illustrated in Fig.\ref{fig:pom-mec}(b),
by using the following Lagrangian
\begin{eqnarray}
L = L_{J/\Psi,\rho^0\pi^0} + L_{\pi NN} + L_{VDM}
\label{eq:larg}
\end{eqnarray}
with
\begin{eqnarray}
L_{J/\Psi,\rho^0\pi^0}&=& -\frac{g_{J/\Psi,\rho^0\pi^0}}{m_{J/\Psi}}
\epsilon^{\mu\nu\alpha\beta}
\partial_\mu {\rho^0}_\nu \partial_\alpha \phi_{J/\Psi,\beta}
 {\phi}_{\pi^0}\,,
\label{eq:L-jrp}\\
L_{\pi NN} &=& -\frac{f_{\pi NN}}{m_\pi}\bar{\psi}_N\gamma_\mu\gamma_5\vec{\tau}\psi_N
\partial^\mu \cdot \vec{\phi}_\pi\,, \\
L_{VDM}&=&\frac{e m_\rho}{f_\rho}A^\mu\rho^0_\mu \,,
\end{eqnarray}
where $\rho^0_\mu$, $\phi_{J/\Psi,\beta}$, $\vec{\phi}_\pi$,
$A^\mu$, and $\psi_N$ are the field operators for $\rho^0$, $J/\Psi$, $\pi$, photon ($\gamma$),
and
nucleon ($N$), respectively. The mass for particle $a$ is denoted as $m_a$. The well determined
coupling constants are $f^2_{\pi NN}/4\pi= 0.079$, $e^2/4\pi=1/137$, and
 $f_\rho=5.33$. To determine $g_{J/\Psi,\rho^0\pi^0}$,
we use $L_{J/\Psi,\rho^0\pi^0}$ given in Eq.(\ref{eq:L-jrp}) to calculate the
decay width
\begin{eqnarray}
\Gamma_{J/\Psi \rightarrow \pi^0\rho^0}&=&(2\pi)\frac{1}{3}
\sum_{\lambda_\rho,\lambda_{J/\Psi}}\int d\Omega_k
|<\vec{k}\lambda_\rho|H|\vec{p}=0,\lambda_{J/\Psi}>|^2
\frac{kE_\pi(k)E_\rho(k)}{m_{J/\Psi}}\,,
\label{eq:width-1}
\end{eqnarray}
where $k$ is defined by $m_{J/\Psi}= E_\pi(k)+E_\rho(k)$, and
\begin{eqnarray}
<\vec{k}\lambda_\rho|H|\vec{\vec{p},\lambda_{J/\Psi}}>
&=&\frac{1}{(2\pi)^{3/2}}\frac{1}{\sqrt{2E_{J/\Psi}(p)}}
\frac{1}{\sqrt{2E_{\rho}(k)}}
\frac{1}{\sqrt{2E_{\pi}(k)}}\nonumber \\
&&\times \epsilon^{\mu\nu\alpha\beta}k^\rho_\mu
\epsilon_{\nu,\lambda_\rho}(k^\rho) p_\alpha
\epsilon_{\beta,\lambda_{J/\Psi}}(p)
[\frac{\Lambda^2_{J/\Psi}}{(\vec{k}^2+\Lambda^2_{J/\Psi})}]^2\,.
\label{eq:width-2}
\end{eqnarray}
Here we have included a dipole cutoff function with a range parameter
$\Lambda_{J/\Psi}$. The four-momenta are defined in the rest frame of $J/\Psi$:
\begin{eqnarray}
p&=&(m_{J/\Psi},\vec{0})\,, \nonumber \\
k^\rho &=& (E_\rho(k),\vec{k}) \,,\nonumber \\
k^\pi &=& (E_\pi(k),-\vec{k}) \,.\nonumber
\end{eqnarray}
By using Eqs.(\ref{eq:width-1})-(\ref{eq:width-2}) and the experimental
value given in Eq.(\ref{eq:jpsi-width}), we find
$g_{J/\Psi,\pi^0\rho^0} = 0.032 $ for a cutoff $\Lambda_{J/\Psi} = 2000$ MeV.

With the Lagrangian Eq.(\ref{eq:larg}), the one-pion-exchange
 invariant amplitude for $\gamma(q)+ N(p) \rightarrow J/\Psi(k)+N(p')$
can  be written as
\begin{eqnarray}
I_{fi} = \bar{u}_{m'_s}(p')\epsilon^*_\mu(k,\lambda_{J/\Psi})M^{\mu\nu}_\pi(p',p)
\epsilon_\nu(q,\lambda_\gamma)u_{m_s}(p)
\label{eq:mex-amp}
\end{eqnarray}
with
\begin{eqnarray}
M^{\mu\nu}_\pi(p',p) &=& G\times F(t)
\frac{1}{t-m^2_\pi} \epsilon^{\mu\nu\alpha\beta}k_\alpha q_\beta
[\gamma \cdot (p'-p)]\,,
\label{eq:opep}
\end{eqnarray}
where $t=(p-p')^2$, and
\begin{eqnarray}
G&=&\frac{e}{f_\rho}\frac{g_{J/\Psi,\rho^0\pi^0}}{m_{J/\Psi}}\frac{f_{\pi NN}}{m_\pi} \,,\\
F(t)&=& F_{\pi NN}(t) F_{J/\Psi,\rho^0\pi^0}(t)\,.
\end{eqnarray}
Here we have introduced a cutoff form factor $F(t)$ to regularize
the interaction verteces. For simplicity, we use the following form
\begin{eqnarray}
F(t)=\left(\frac{\Lambda^2}{\Lambda^2-t}\right)^n\,.
\label{eq:mex-ff}
\end{eqnarray}
We set $n=4$, and $\Lambda = \Lambda_{J/\Psi} = 2000$ MeV.

\section{Photo-production of $[B]_{J/\Psi}$ bound state}
\subsection{Reaction Mechanism}
With the impulse approximation, we assume that a $J/\Psi$ is produced on a nucleon in the target nucleus
$A$ and then is attracted by  a spectator system $B$ to form a bound state $[B]_{J/\Psi}$. For simplicity, $[B]_{J/\Psi}$ is denoted as $d$ in the following formula.
\begin{figure}[b]
\centering
\epsfig{file=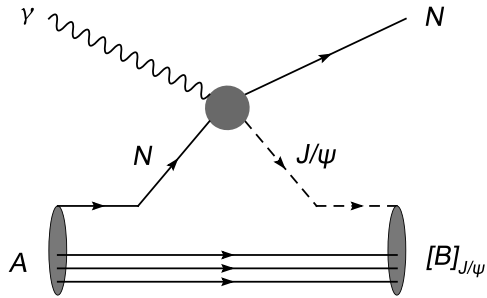, width=0.4\hsize}
\caption{The impulse approximation mechanism of
$\gamma + A \rightarrow N + [B]_{J/\Psi}$ reaction.
$A$ is a nucleus with mass number $A$ and $B$ could be a nucleus with mass
number $(A-1)$ or a $[q^{3(A-1)}]$ multi-quark cluster.
}
\label{fig:ghe4}
\end{figure}

With the mechanism illustrated in Fig.\ref{fig:ghe4}, the cross section of $\gamma(\vec{q})+ A(-\vec{q}) \rightarrow N(\vec{p}) + d(-\vec{p})$ in the center of mass system can  be written as
\begin{eqnarray}
\frac{d\sigma}{d\Omega} &=&[\frac{(2\pi)^4E_A(q)}{W}]
[\frac{pE_N(p)E_d(p)}{W}] \nonumber \\
&&\times \frac{1}{2}\frac{1}{2J_A+1}\sum_{\lambda, M_{J_A}}\sum_{m_s,m_d}
|<\vec{p}m_s, \Psi_{j_dM_d}|T(W)|\vec{q}\lambda, \Phi^A_{J_A,M_{J_A}}>|^2\,,
\label{eq:d-prod-1}
\end{eqnarray}
where
\begin{eqnarray}
&& <\vec{p}m_s, \Psi_{j_dM_d}|T(W)|\vec{q}\lambda, \Phi^A_{J_A,M_{J_A}}>
=\sum_{j,m_j}\sum_{j_\alpha,m_{j_\alpha}}
< \Psi_{j_dM_d}|a^\dagger_{jm_j}b_{j_\alpha,m_{j_\alpha}}|\Phi^A_{J_A,M_{J_A}}>
\nonumber \\
&&\times [\sum_{m_{J/\Psi},m_{s_\alpha}}
\int d\vec{k}
\chi^*_{j,m_j}(\vec{Q}_d,m_{J/\Psi})
<\vec{p}m_s, \vec{k}m_{J/\Psi}|t(W)|\vec{q}\lambda,\vec{p}_\alpha m_{s_\alpha}>
\phi_{j_\alpha,m_{j_\alpha}}(\vec{Q}_A,m_{s_\alpha})\,.
\nonumber \\
\label{eq:d-prod-2}
\end{eqnarray}
Here $a^\dagger_{jm_j}$ is the creation operator for a $J/\Psi$ with wavefunction $\chi_{j,m_j}(\vec{Q}_d,m_{J/\Psi})$, $b_{j_\alpha,m_{j_\alpha}}$ an annihilation operator for a nucleon with
wavefunction $\phi_{j_\alpha,m_\alpha}(\vec{Q}_A,m_{s_\alpha})$, and the $\gamma + N \rightarrow J/\Psi + N$ amplitude is
\begin{eqnarray}
 <\vec{p}m_s, \vec{k}m_{J/\Psi}|t(W)|\vec{q}\lambda,\vec{p}_\alpha m_{s_\alpha}> &=& \nonumber
 \frac{1}{(2\pi)^3} \frac{1}{\sqrt{2E_{J/\Psi}(k)}}
\sqrt{\frac{m_N}{E_N(p_\alpha)}}\frac{1}{\sqrt{2q}}
\sqrt{\frac{m_N}{E_N(p)}} \nonumber \\
&&\times [\bar{u}_{m_s}(p)\epsilon^*_{\mu}(k,m_{J/\Psi})\{
M^{\mu\nu}_\pi(p,p_\alpha)
+\mathcal{M}^{\mu\nu}_\mathbb{P}(p,p_\alpha)\}\nonumber \\
&&\,\,\times \epsilon_{\nu}(q,\lambda)u_{m_{s_\alpha}}(p_\alpha))]\,,
\nonumber \\
\label{eq:t-mx}
\end{eqnarray}
where $M^{\mu\nu}_\pi$ is the pion-exchange amplitude given in Eq.(\ref{eq:opep}), and $\mathcal{M}^{\mu\nu}_\mathbb{P}$ is the Pomeron-exchange amplitude in Eq.(\ref{eq:MP}).

For simplicity, we will only perform calculations for the reactions on $^3He$ and $^4He$.
For estimations of cross sections on these target nuclei,
it is sufficient to use the s-wave harmonic oscillator wavefunctions
for both the target $A$ and $B$ in the
$[B]_{J/\Psi}$ bound state. We also only consider the case that the $J/\Psi$ in the produced bound $B_{J/\Psi}$ is  on an s-wave orbital. For the case  $B=A-1$ nuclear system, we thus write the initial ($|\Phi^A>$) and final ($|\Psi>$) nuclear states as
\begin{eqnarray}
 |\Phi^A>  &=&[ |N> \otimes |\Phi^{A-1}> ]_{L=0} \,,\label{eq:a-wf}\\
|\Psi>  &=& [|J/\Psi> \otimes |\Phi^{A-1}>]_{L=0}\,, \label{eq:j-wf}
\end{eqnarray}
where $L$ is the relative angular momentum between $N$ or $J/\Psi$ and the $(A-1)$ nucleus. Explicitly, we have
\begin{eqnarray}
|\Phi^A_{J_A,M_{J_A}}> &=& \sum_{M_{J_{A-1}}}\sum_{j_\alpha, m_{j_\alpha}}
<J_AM_{J_A}|j_\alpha J_{A-1}m_{j_\alpha} M_{J_{A-1}} >
b^\dagger_{j_\alpha m_{j_\alpha}} |\Phi^{A-1}_{J_{A-1},M_{J_{A-1}}}>\,,
\nonumber \\
|\Psi_{J_d,M_{J_d}}> &=& \sum_{M_{J_{A-1}}}\sum_{j, m_j}
<J_dM_{d}|jJ_{A-1} m_j M_{J_{A-1}} >
a^\dagger_{j m_j} |\Phi^{A-1}_{J_{A-1},M_{J_{A-1}}}>\,.
\end{eqnarray}
Then the momentum variables in Eqs.(\ref{eq:d-prod-2}) and (\ref{eq:t-mx}) are
\begin{eqnarray}
\vec{p}_\alpha &=&\vec{p}+\vec{k}-\vec{q}\,, \\
\vec{p}_\beta &=& -\vec{p}-\vec{k} =-\vec{q}-\vec{p}_\alpha\,, \\
\vec{Q}_A&=&\frac{\vec{p}_\alpha E_{A-1}(\vec{p}_\beta)-
\vec{p}_\beta E_N(\vec{p}_\alpha)}
{E_{A-1}(\vec{p}_\beta) +E_N(\vec{p}_\alpha)}\,, \\
\vec{Q}_d&=&\frac{\vec{k} E_{A-1}(\vec{p}_\beta)
-\vec{p}_\beta E_{J/\Psi}(\vec{k})}
{E_{J/\Psi}(\vec{k}) +E_{A-1}(\vec{p}_\beta)}\,,
\label{eq:d-momentum}
\end{eqnarray}
where $\vec{Q}_d$ ($\vec{Q}_A$) is the relativistic relative momentum between
$J/\Psi$ ($N)$ and the $(A-1)$ nuclear system.

For the target $^4He$, we have $J_A=0$ and assume that $|\Phi^{A-1}_{J_{A-1},M_{J_{A-1}}}>$
is the $^3He$ ground state with $J_{A-1}=1/2$. We then have the following
simplicities:
\begin{eqnarray}
< \Psi_{j_dM_d}|a^\dagger_{jm_j}b_{j_\alpha,m_{j_\alpha}}|\Phi^A_{J_A,M_{J_A}}>
&=&<J_AM_{J_A}|j_\alpha J_{A-1} m_{j_\alpha}M_{J_{A-1}}>
<J_dM_{d}|jJ_{A-1} m_j M_{J_{A-1}} > \nonumber \\
&\rightarrow& \frac{1}{\sqrt{2}}
<J_dM_{J_d}| j_\alpha j - m_{j_\alpha} m_j>\,,
\end{eqnarray}
and
\begin{eqnarray}
\chi_{j,m_j}(\vec{Q}_d,m_{J/\Psi}) &=& \delta_{j,1}\delta_{m_j,m_{J/\Psi}}
\frac{1}{\sqrt{4\pi}}F({Q}_d) \label{eq:wf-jpsi-0}\,,\\
\phi_{j_\alpha,m_{j_\alpha}}(\vec{Q}_A,m_{s_\alpha})&=&\delta_{j_\alpha,1/2}
\delta{m_{j_\alpha},m_{s_\alpha}}\frac{1}{\sqrt{4\pi}} R({Q}_A)\,,
\end{eqnarray}
Eq.(\ref{eq:d-prod-2}) then becomes
\begin{eqnarray}
&& <\vec{p}m_s, \Psi_{j_dM_d}|T(W)|\vec{q}\lambda, \Phi^A_{J_A,M_{J_A}}>
\nonumber \\
&&=\sum_{M_{J_{A-1}}}\sum_{m_{J/\Psi},m_{s_\alpha}}
<J_AM_{J_A}|j_\alpha J_{A-1} m_{j_\alpha}M_{J_{A-1}}>
<J_dM_{d}|jJ_{A-1} m_j M_{J_{A-1}}> \nonumber \\
&&\times [\int d\vec{k}
\frac{1}{\sqrt{4\pi}}F({Q}_d)
<\vec{p}m_s \vec{k}m_{J/\Psi}|t(W)|\vec{q}\lambda,\vec{p}_\alpha m_{s_\alpha}>
\frac{1}{\sqrt{4\pi}} R({Q}_A)]\,.
\label{eq:d-prod-3}
\end{eqnarray}

We have applied the formula Eqs.(\ref{eq:d-prod-1}) and (\ref{eq:d-prod-3}) to estimate the production cross section on $^4He$. We use the usual s-wave harmonic oscillator wavefunction with $b=1.32$ fm for the target $^4He$
\begin{eqnarray}
R(p)=
[Ne^{\frac{-b^2p^2}{2}}]
\end{eqnarray}
with the normalization
$\int R^2(p) p^2 dp=1$.
For the $J/\Psi$ wavefunction in $d = [^3He]_{J/\Psi}$, we will generate a s-wave $\psi_{J/\Psi}(r)$ from a potential $V_{J/\Psi,B}(r)$ with the normalization $\int r^2 dr |\psi_{J/\Psi}(r)|^2=1$. The wavefunction in Eq.(\ref{eq:wf-jpsi-0}) and also in (\ref{eq:d-prod-3}) can then be calculated from
\begin{eqnarray}
F(p) &=& \int_0^\infty r^2 dr j_0(pr) \psi_{J/\Psi}(r)\,, \label{eq:jpsi-tra}
\end{eqnarray}
where $j_0(z)$ is the spherical Bessel function.
The form of  $V_{J/\Psi,B}(r)$ will be discussed in section IV.

The above formula can be easily extended to investigate other possible
impulse approximation
mechanisms as far as all wavefucnctions  in the bound $A$ and $[B]_{J/\psi}$ are all in s waves. This is what we will need in section IV when we consider the production of $[q^6]_{J/\Psi}$ from the $q^6$-N component of $^3He$.

\section{Results}

\subsection{Models of $\gamma+ N \rightarrow J/\Psi +N$ reaction}

\begin{figure}[htbp] \vspace{-0.cm}
\begin{center}
\includegraphics[width=0.49\columnwidth]{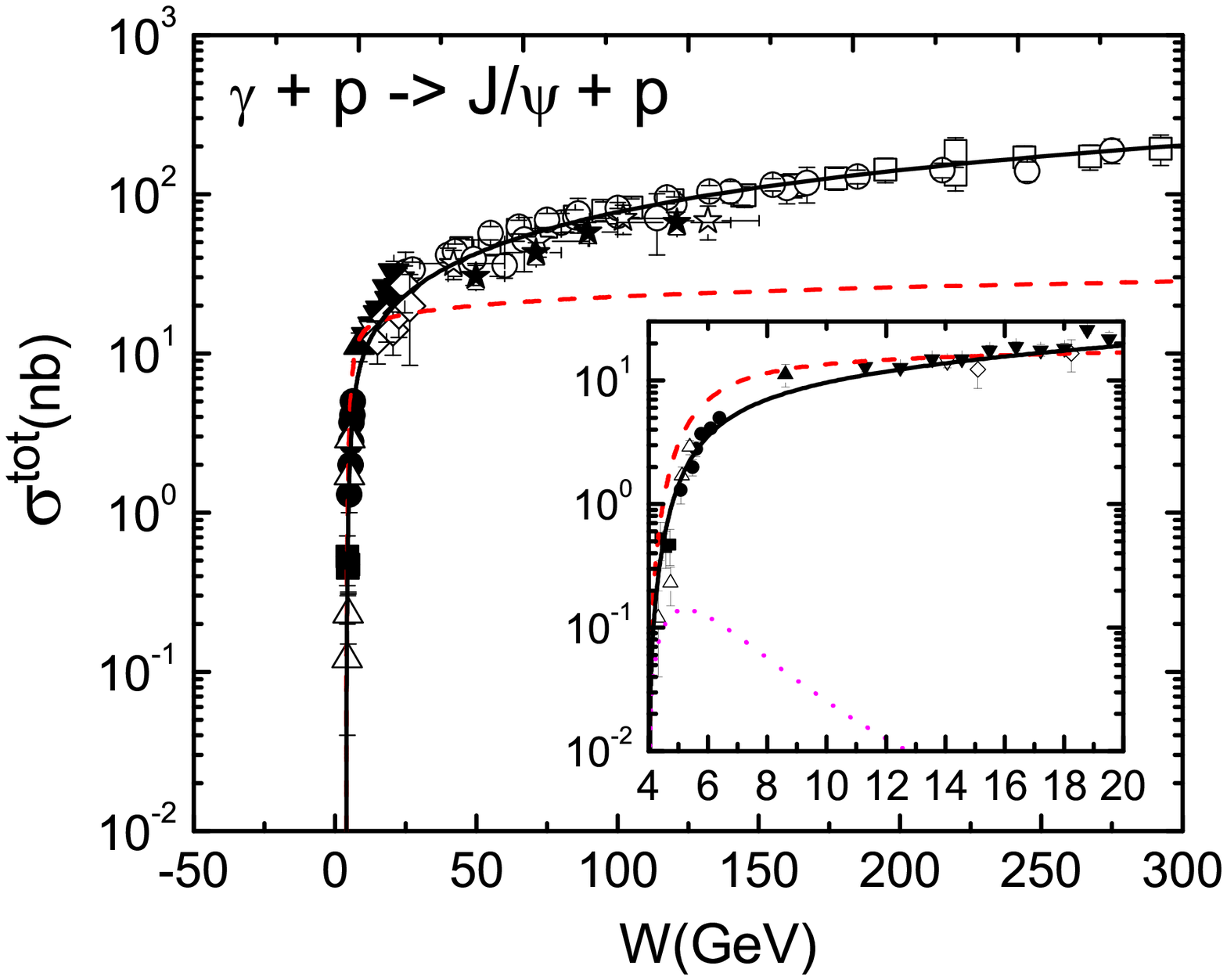}
\includegraphics[width=0.49\columnwidth]{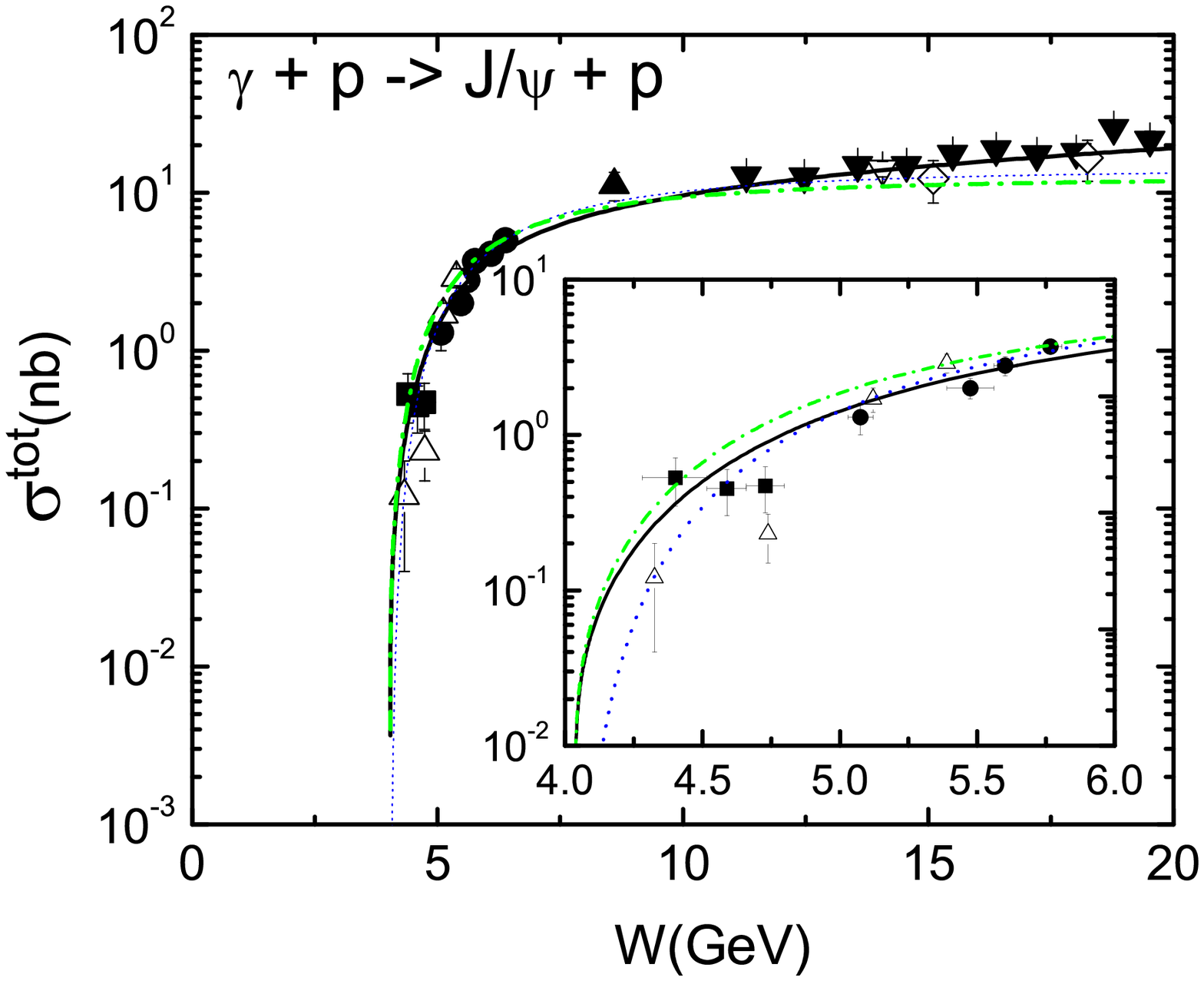}
\caption{The total cross section of the $\gamma + p \to J/\psi + p$ reaction
as function of  the
$\gamma p$  invariant mass $W$.
Left: the black solid
and red dashed are the results from
using  ($\alpha_0=1.25$,  $\beta_c=0.84$  GeV$^{-1}$; called the PM model)
and ($\alpha_0=1.08$, $\beta_c=1.21$  GeV$^{-1}$) within the
model which
 include both
the $\pi$-exchange and Pomeron-exchange mechanisms.
The  magenta dotted curve in the insert
is the contribution from the $\pi$ exchange.
Right: the black solid, blue dotted and green dot-dashed lines are
from the PM model, the $2g$ model of Ref.\cite{brodsky-2},
and the $2g+3g$ model based on Eq.(\ref{tmatrix}).
The experimental data are from \cite{Barate:1986fq,Binkley:1981kv,Camerini:1975cy,Denby:1983az,Frabetti:1993ux,Gittelman:1975ix,Anderson,H1,ZEUS}.} \label{fg:jpsiphoto}

\end{center}
\end{figure}

We first develop a model consisting of
Pomeron-exchange and pion-exchange mechanisms, as described in section II. With the parameters specified there, we try to fit the available total cross section data of $\gamma+ p \rightarrow J/\Psi+p$  up to invariant mass
$W = 300$ GeV by only adjusting the charmed quark-Pomeron coupling constant $\beta_c$. With $\beta_c =1.21$ we only able to fit the data up to 20 GeV. Clearly, the result at high energy is not satisfactory as shown in the red dashed curve in the left-hand side of Fig.\ref{fg:jpsiphoto}. We then find that by changing $\alpha_0$ of the Regge trajectory in the Pomeron propagator Eq.(\ref{eq:regge-g}) from $\alpha_0=1.08$, as determined in the previous fits\cite{OL02} to
 the total cross sections of
$\rho$ and $\omega$ photo-production, to $\alpha_0=1.25$, we are able to get a very good fit to the data by choosing $\beta_c = 0.84$  GeV$^{-1}$.
Our fit is the solid black curve in the left-hand side of Fig.\ref{fg:jpsiphoto}. We thus will use the model with $\alpha_0=1.25$ and $\beta_c = 0.84$  GeV$^{-1}$
(PM  model) in our investigations. As also seen in the insert in the left-hand side of Fig.\ref{fg:jpsiphoto}, the contribution
(magenta dotted curve) from the pion-exchange amplitude,
 as defined by Eqs.(\ref{eq:mex-amp})-(\ref{eq:mex-ff}), is very weak except in the very near threshold region.

We next consider the model of Ref.\cite{brodsky-2} based on the two-gluon ($2g$)
and three-gluon ($3g$) exchange
mechanisms. In terms of the normalization defined by Eq.(\ref{eq:tot-amp}),
the amplitude of this model can be written as
 \begin{eqnarray}
<\vec{k}\lambda_{J/\psi}m'_s|t(W)|\vec{q}\lambda_{\gamma}m_s>
&=&\frac{1}{(2\pi)^3}\frac{1}{\sqrt{2E_{J/\psi}(k)}}\sqrt{\frac{m_N}{E_{N}(k)}}
\sqrt{\frac{m_N}{E_{N}(q)}}\frac{1}{\sqrt{2q}} \nonumber \\
&&\times
\frac{4\sqrt{\pi}}{\sqrt{6}}\frac{qw}{m_N}[\textbf{M}_{2g}+\textbf{M}_{3g}]
 \label{tmatrix}
\end{eqnarray}
with
\begin{eqnarray}
\textbf{M}_{2g}&=&\frac{A_{2g}}{4\sqrt{\pi}}\frac{1-x}{Rm_{J/\psi}}e^{bt/2}\,,\\\label{brodsky2g}
\textbf{M}_{3g}&=&\frac{A_{3g}}{4\sqrt{\pi}}\frac{1}{R^2m^2_{J/\psi}}e^{bt/2}\,,\\\label{brodsky3g}
x&=&\frac{2m_Nm_{J/\psi}+m^2_{J/\psi}}{W^2-m^2_p}\label{brodskyx}\,.
\end{eqnarray}
where  $R=1$ fm, $b=1.13$ GeV$^{-2}$  are taken from Ref. \cite{brodsky-2}.
 We follow Ref.\cite{brodsky-2}
to determine the parameters $A_{2g}$ and $A_{3g}$ by fitting
the data up to only 20 GeV.
In the two-gluon-exchange model ($2g$), we set $A_{3g}=0$ and obtain
$A_{2g}=0.028$ MeV$^{-2}$ from the fit.
In the $2g+3g$ model, the fit is obtained by choosing
 $A_{2g}=0.023$ MeV$^{-2}$ and $A_{3g}=2000$ MeV$^{-2}$. The fits for the  $2g$ and $2g+3g$ models are the dotted and
dot-dashed curves in the right-hand side of Fig.\ref{fg:jpsiphoto}, respectively. Clearly, they have differences with that (black solid) of the PM model, as
can be seen more clearly  in the insert
 in the right-hand side of Fig.\ref{fg:jpsiphoto}.
Here we also see that
the data in the region near the $J/\Psi$ production threshold are
very limited and uncertain.
We  will therefore  perform calculations using the PM, $2g$, and $2g+3g$ models
 to examine the model dependence of our predictions. Clearly,
precise data in the near threshold region are needed to make progress.

\subsection{Photo-production of $J/\Psi$-Nucleus bound states}

Following the previous investigations \cite{brodsky90,gao01}, we assume that
the interaction between a $J/\Psi$ and a nucleus with mass number $A$
can be parameterized as a non-relativistic potential of the following Yukawa form
\begin{eqnarray}
V_{J/\Psi, A}(r) = -\alpha_A\frac{e^{-\mu_A r}}{r} \,.
\label{eq:yukawa-n}
 \end{eqnarray}
There exists two different approaches to determine the parameters
$ \alpha_A$ and  $\mu_A$ for the nucleon with $A=1$.
We will explain these in the following two subsections.

\subsubsection{Pomeron-quark coupling model}
Motivated by the previous studies in Quantum Electrodynamics, it is assumed
in the approach of Ref.\cite{brodsky90}
that the $J/\Psi$-A forward angle scattering amplitude at very high energy can be related to the matrix element of the potential Eq.(\ref{eq:yukawa-n}) which is understood to be  valid only in the region where $J/\Psi$ moves
non-relativistically.
They further assume that the $J/\Psi$-A amplitudes can be
 calculated by using the Pomeron-exchange
model of Dannachie and Landshoff\cite{donn84}. In the very
high energy approximation,
the differential cross section of $J/\Psi$-A
elastic scattering can be related to the
parameters $\alpha_A$ and $\mu_A$ of the potential Eq.(\ref{eq:yukawa-n}) by
 following relation
\begin{eqnarray}
\frac{d\sigma}{dt} (J/\Psi\, A\rightarrow J/\Psi\, A) &=&
\frac{[2\beta_c F_{J/\Psi}(t)]^2
[3A\beta_{u/d} F_{A}(t)]^2}{4\pi} \label{eq:dl-jpsi-1} \\
&=& \frac{4\pi\alpha^2_A}{(-t+\mu^2_A)^2}\,,
\label{eq:dl-jpsi-2}
\end{eqnarray}
where $t$ is the momentum-transfer squared, $\beta_{u/d}$ ($\beta_c$) is the Pomeron coupling with the $up/down$ ( $charmed$ ) quarks, $F_{J/\Psi}(t)$ and $F_{A}(t)$ are the form factors for $J/\Psi$ and the nucleus with mass number $A$, respectively. They further
assume that in the  $t\rightarrow 0$ limit, the slope of
${d\sigma}/{dt} $ is mainly determined by $dF_{A}(t)/dt$ and that
 $F_{A}(t)$ can be identified with the nuclear electromagnetic form factor. One then gets the following relations
\begin{eqnarray}
\mu^{-2}_A&=& |\frac{dF_A(t)}{dt}|_{t=0} = \frac{<R^2_A>}{6}\,, \\
\alpha_A &=& \frac{[2\beta_{c}][3A\beta_{u/d}]}{4\pi}\mu^2_A\,.
\label{eq:alpha}
\end{eqnarray}
The radius $<R^2_A>^{1/2}$ can be taken from Ref.\cite{nucl-data}. The Pomeron-quark coupling constants can be taken from fits to the data of meson-nucleon scattering or  photo-production of vector mesons. Once $\alpha_A$ and $\mu_A$ of the potential Eq.(\ref{eq:yukawa-n}) are determined, we can predict the possible $J/\Psi$-nucleus bound states.
In Table \ref{tab:bound-1}, we list our results for proton ($A=1)$, $^3He$ (A=3) and $^{12}C$ (A=12) for various sets of Pomeron-quark coupling constants. The first rows in the results for each $A$ are based on the flavor independent
$\beta_{u/d}=\beta_{c} = 1.85$ GeV$^{-1}$ of Ref.\cite{brodsky90}. The other two results  use the Pomeron-quark coupling constants $\beta_{u/d}=2.05$ GeV$^{-1}$ determined\cite{OL02} in the fits to the data of photo-production of $\rho$ and $\omega$, and $\beta_c$ determined from the fits described in section IV.A.

\begin{table}[t]
\begin{center}
\caption{Parameters for determining the potential
Eq.(\ref{eq:yukawa-n}) using the Pomeron-quark coupling model defined
by Eqs.(\ref{eq:dl-jpsi-2})-(\ref{eq:alpha}). The predicted binding energies
(B.E.) for proton ($A=1)$,
$^3He$ (A=3) and $^{12}C$ (A=12) are also listed.  } \label{tab:bound-1}
\begin{tabular}{ccccccc}\hline
 A & $<R^2_A>^{1/2}$   &  $\mu_A$ & $\beta_{u/d}$ & $\beta_c$    & $\alpha_A$ & B.E. \\
   & (GeV$^{-1}$)      &   (GeV)  & (GeV$^{-1}$)  & (GeV$^{-1}$) &            & (MeV)\\
 1 &    3.9            &  0.63   &    1.85       &  1.85        & 0.64       &   -  \\
   &                   &          &    2.05       &  1.21        & 0.47       &   -   \\
   &                   &          &    2.05       &  0.84        & 0.33       &   -   \\
\hline
3  &    9.5            &  0.26   &    1.85       &  1.85        & 0.33       & 19.86   \\
   &                   &          &    2.05       &  1.21        & 0.23       & 3.27   \\
   &                   &          &    2.05       &  0.84        & 0.16       & 0.04 \\
\hline
12 &   12.69           &  0.19   &    1.85       &  1.85        & 0.73       & 280.0  \\
   &                   &          &    2.05       &  1.21        & 0.53       & 165.0   \\
   &                   &          &    2.05       &  0.84        & 0.37       & 67.0 \\
\hline
\end{tabular}
\end{center}
\end{table}

\begin{figure}[htbp] \vspace{-0.cm}
\begin{center}
\includegraphics[width=0.49\columnwidth]{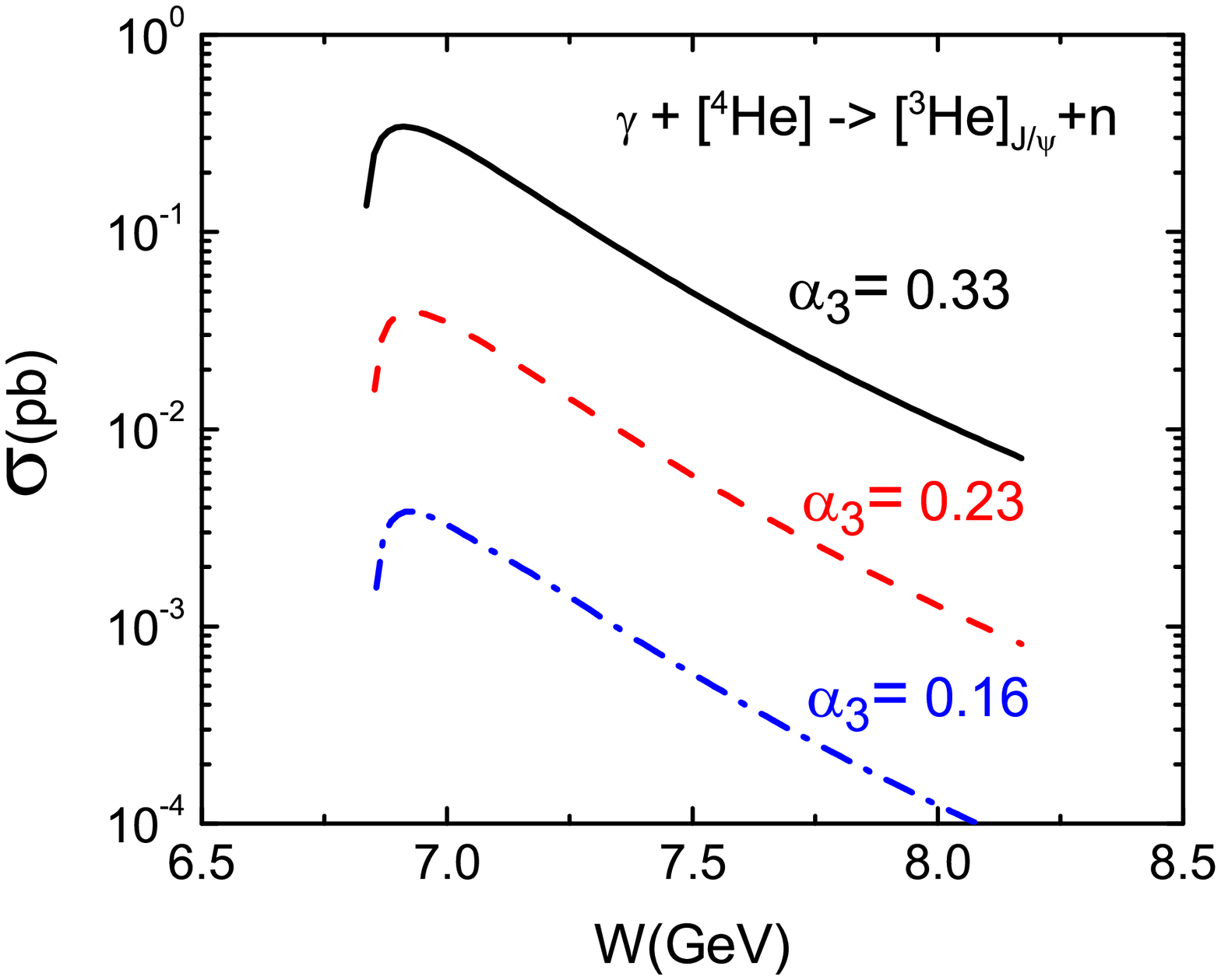}
\includegraphics[width=0.49\columnwidth]{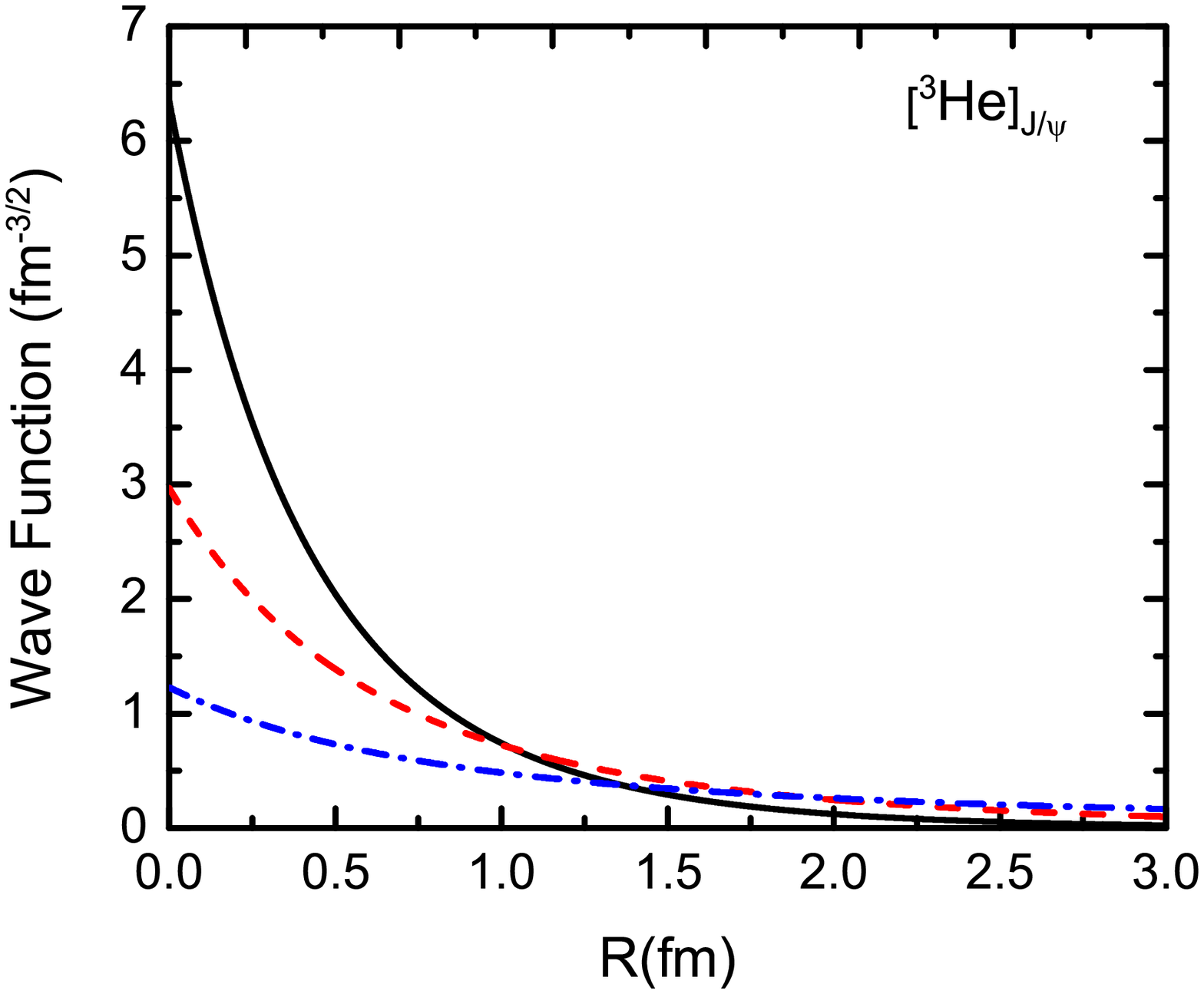}
\caption{The total cross section of
$\gamma + [^4He] \to ^3He_{[J/\psi]} + n$ as function of $\gamma$-$^4He$
invariant mass $W$ (left) and
 the wave function (right) for $J/\psi - ^3He$ system. The black solid, red
dashed and blue dotted-dashed lines are calculated by using
the potential Eq.(\ref{eq:yukawa-n}) with $A=3$,
 $\mu_A=0.257$ GeV and $\alpha_A =0.33, 0.23,0.16$, respectively.} \label{fig:compare}
\end{center}
\end{figure}

\begin{figure}[htbp] \vspace{-0.cm}
\begin{center}
\includegraphics[width=0.49\columnwidth]{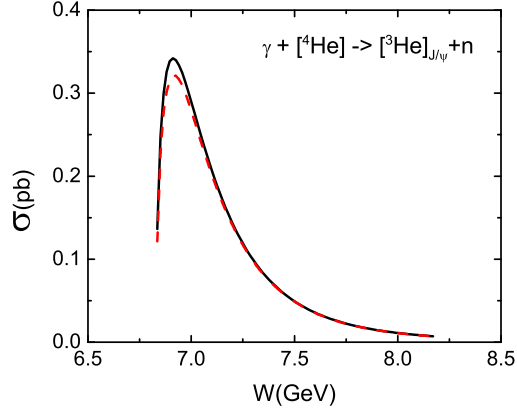}
\caption{The cross section of $\gamma + [^4He] \to [^3He]_{[J/\psi]} + n$
as function of $\gamma$-$^4He$
invariant mass $W$. The red dashed curve is
obtained from keeping only the contribution
from  the $J/\Psi$ wavefunction  with
$ k  > 1400$ MeV in the integration of Eq.(\ref{eq:d-prod-3}).}
\label{fig:mainHe4}
\end{center}
\end{figure}

With the determined potential parameters $\alpha_A$ and
$\mu_A$, the predicted binding energies ($B.E.$)
 for each considered nuclear system are listed in the last column
of Table \ref{tab:bound-1}.
For the $A=1$ case, we see that there is no $J/\Psi$-$N$ bound state.
 But all three models predict bound $[^3He]_{J/\Psi}$ and $[^{12}C]_{J/\Psi}$ states. In the left-hand side of Fig.\ref{fig:compare},
we show the predicted cross sections of $\gamma +^4He \to [^3He]_{J/\Psi} + n$. Clearly, the predicted cross sections depend on the Pomeron-quark coupling constants. Furthermore, their magnitudes depend sensitively on the binding energy (B.E.) of the predicted $[^3He]_{J/\Psi}$ system. As the binding energy decreases from 19.86 MeV to 0.04 MeV, the predicted cross sections drop by two orders in magnitude. This can be understood from the right-hand side
of Fig.\ref{fig:compare} where we compare the $J/\Psi$-$^3He$ relative wavefunctions which are used in predicting the
cross sections in the left-hand side.
We see that the wavefunction (solid black) for B.E. $= 19.86 $ MeV is much shorter range than the other two cases and hence gives more cross sections in this large momentum-transfer reaction.
This is explicitly illustrated in Fig.\ref{fig:mainHe4} where we show that
 the cross section (red dashed curve)
calculated from keeping only
the high momentum part ($p_{J/\psi} > 1400$ MeV) of the $J/\Psi$ wavefnction
in the integration in Eq.(\ref{eq:d-prod-3}) is very close to the full calculation (solid black curve).

\begin{figure}[htbp] \vspace{-0.cm}
\begin{center}
\includegraphics[width=0.49\columnwidth]{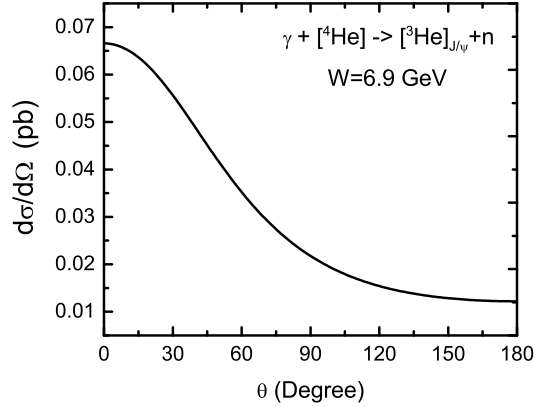}
\caption{The differential cross section of $\gamma + [^4He] \to [^3He]_{[J/\psi]} + n$ vs the angle of out going $N$ with the center mass $6.9$ GeV.}
\label{fig:diffHe4}
\end{center}
\end{figure}

\begin{figure}[htbp] \vspace{-0.cm}
\begin{center}
\includegraphics[width=0.49\columnwidth]{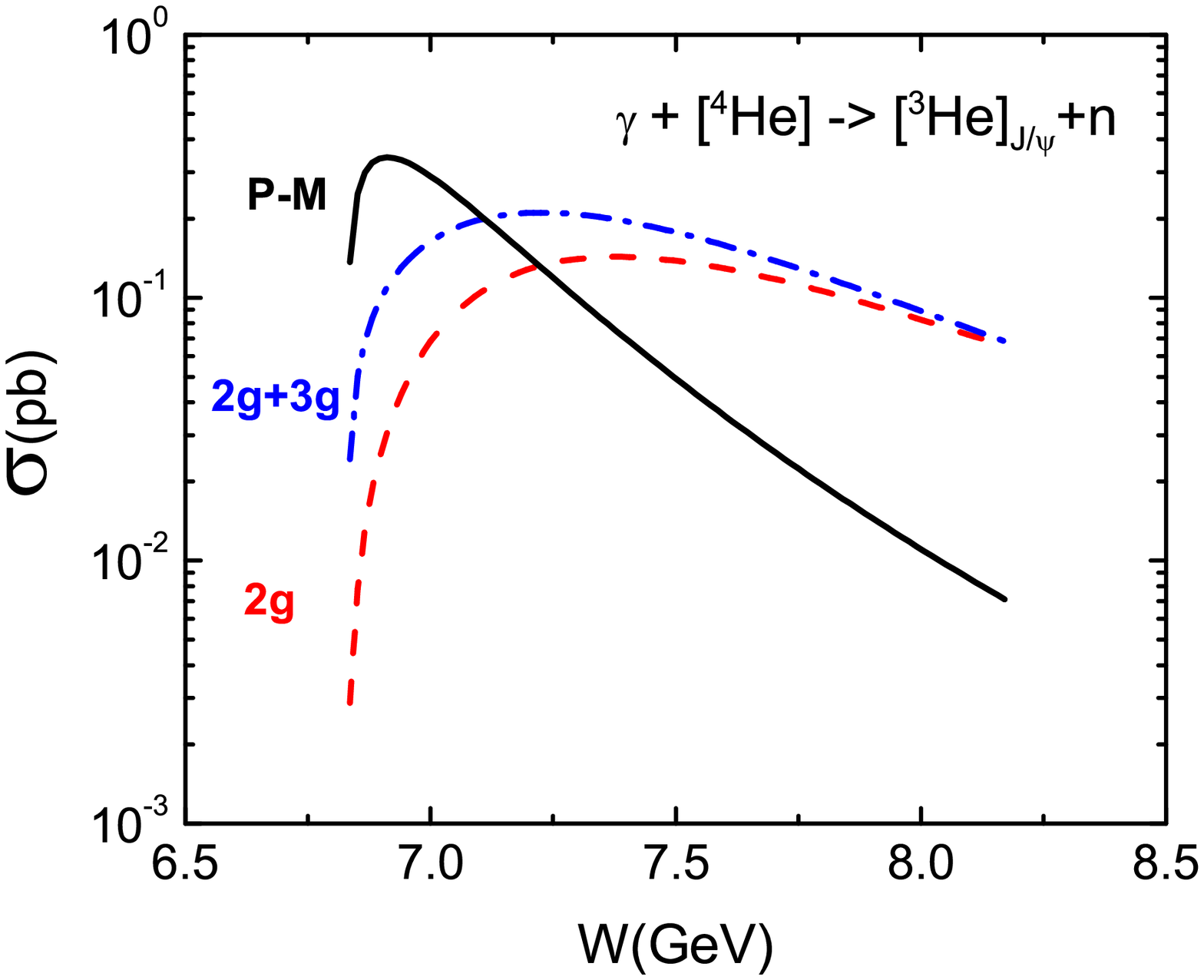}
\caption{The total cross section of $\gamma + [^4He] \to [^3He]_{[J/\psi]} + n$ vs the certain mass of system. The black solid, red dashed and blue dot-dashed lines are for the pomeron and $\pi$ exchange, Brodsky's 2g model and 2g+3g model, respectively.}
\label{fig:mechHe4}
\end{center}
\end{figure}

In Fig.\ref{fig:diffHe4}, we see that the predicted differential cross sections are forward peaked, as expected from the Pomeron-exchange mechanism.
In Fig.\ref{fig:mechHe4}, we show that the predicted cross sections depend on the $\gamma + N \rightarrow J/\Psi +N$ model. Their maximum values are,
 however, comparable $\sim 0.1 - 0.3 $ pico-barn. Clearly, it is important to get
accurate data of $\gamma + N \rightarrow J/\Psi +N$ at low energies
to refine
the employed model
for making more precise predictions.

The $[^{12}C]_{J/\Psi}$ can be produced by $\gamma +^{13}C \rightarrow [^{12}C]_{J/\Psi} + n$.
However, making predictions for the cross sections of this process is beyond the scope of this paper
since the simple s-wave description of the nuclei in section III is no longer a
reasonable approximation for nuclei heavier than $^4He$.

\subsubsection{Folding model}

While all three $J/\Psi$-$N$ models listed in Table \ref{tab:bound-1} do not have bound states, there exist a possibility that adding the $J/\Psi$-$N$ interactions from the nucleons in a nucleus could lead to bound states.
To explore this possibility,
we follow the usual nuclear physics approach to construct
 a folding potential for the interaction between a $J/\Psi$ and a nuclear
system,
\begin{eqnarray}
V_{J/\Psi, A}(r) =\int v_{J/\Psi,N}(\vec{r}-\vec{r}^{\,\,'})
\rho_A(\vec{r}^{\,\,'})d \vec{r}^{\,\,'}\,,
\label{eq:potfol}
\end{eqnarray}
where $v_{J/\Psi,N}(r)=V_{J/\Psi, 1}(r)$ as defined by Eq.(\ref{eq:yukawa-n})
with $A=1$, and the nuclear density is normalized by
\begin{eqnarray}
\int
\rho_A(\vec{r}^{\,\,'})d \vec{r}^{\,\,'}=A\,.
\end{eqnarray}
For $^3He$ we use $ \rho_A(\vec{r}) = \rho_0 e^{-r^2/b^2}$ with
$b = 1.32$ fm which is obtained by fitting
the $^3He$ charge form factor at low momentum-transfer.
For heavy nuclei, we use the Woods-Saxon form\cite{bohr-moltt}
\begin{eqnarray}
\rho_A(\vec{r}) = \rho_0 \frac{1}{1+e^{(r-R)/t}}
\end{eqnarray}
with $R = 1.1 A^{-1/3}$ fm and $t=0.53$ fm.

Our results using the parameters of Ref.\cite{brodsky90} to
calculate $v_{J/\Psi,N}(\vec{R})$ in Eq.(\ref{eq:potfol})
are listed in the first row of Table \ref{tab:bound-2}. We see that the folding model
gives $1.62$ MeV ($7.0$ MeV) for  $[^3He]_{J/\Psi}$ ( $[^{12}C]_{J/\Psi}$)
which are much less than $19.86$ MeV ( $280$ MeV)
 listed in Table \ref{tab:bound-1}. The predicted cross sections for $\gamma +^4He \rightarrow [^3He]_{J/\Psi} + n$ are also found to be much weaker, close to the
blue dot-dashed curve ($\alpha_3=0.16$)
in   Fig.\ref{fig:compare}.
Clearly, it is difficult to measure such a loosely bound $[^3He]_{J/\Psi}$ state.

To examine the model dependence, we also consider folding potentials by using
three other $J/\Psi$-$N$ models. Two\cite{brodsky-1,russia} of them are
constructed by using the results from the heavy quark effective field theory calculation  by Peskin\cite{pesk79}. The third one\cite{lqcd} is from Lattice QCD calculation. Their results can also be written in the Yukawa form of Eq.(\ref{eq:yukawa-n}) with $A=1$. We find that these three models do not generate a $[^3He]_{J/\Psi}$ bound state as indicated in Table \ref{tab:bound-2}. For $^{12}C$, the binding energies
from folding model are much weaker than those listed in
Table \ref{tab:bound-1} from the Pomeron-quark coupling model.

\begin{table}[ht]
\begin{center}
\caption{The binding energy of  $J/\psi$-nucleus
 calculated with the folding potential defined by Eq.(\ref{eq:potfol})
 with parameters of $v_{J/\Psi,N} = -\alpha_1\frac{e^{-\mu_1r}}{r}$ taken
from different references. The parameters of Ref.\cite{brodsky-1} are
obtained from  reproducing the scattering length $a=-0.24$ fm given in
Ref.\cite{brodsky-1}. (A Gaussian form of the $v_{J/\Psi,N}$ was used in
Ref.\cite{brodsky-1})}
\label{tab:bound-2}
\begin{tabular}{cccccc}\hline
{Model}&\multicolumn{2}{c}{Parameter(MeV)} &   \multicolumn{3}{c}{Binding Energy(MeV)}\\
 & $\alpha_1$  &  $\mu_1$ (GeV)   &  $[H]_{J/\Psi}$ & $[^3He]_{J/\Psi}$ &
$[^{12}C]_{J/\Psi}$   \\
&&&&&\\
Ref.\cite{brodsky90}& 0.64        &   0.63            &  -   & 1.62  &   7.0           \\
Ref.\cite{brodsky-1}& 0.20        &   0.63           &  -   &   -    &   0.91         \\
Ref.\cite{lqcd}& 0.10        &   0.63            &  -   &   -    &   0.003        \\
Ref.\cite{russia}& 0.06        &   0.63            &  -   &   -    &   -             \\
\hline
\end{tabular}
\end{center}
\end{table}

\subsection{Photo-production $J/\Psi$-$(q^6)$ bound states}
In Ref.\cite{brodsky-2}, it was suggested that a $c\bar{c}$ system could
interact strongly with the color octet 3-quark $[\bar{q}^3]_8$ component of the six-quark cluster ($[q^6]=[q^3]_8[\bar{q}^3]_8$) which could dominant the short-range part of the
deuteron wavefunction. The possible attractive force between a $J/\Psi$ and a six-quark cluster was
 suggested in the study of Ref.\cite{brodsky-3} where the excitation of a hidden charm $|qqqqqqc\bar{c}>$
state is introduced to explain the spin correlation of $pp$ elastic scattering near the $J/\Psi$ production threshold. Here we examine the condition under which a bound  $[q^6]_{J/\Psi}$ color singlet state can be produced in the $\gamma + ^3He \rightarrow [q^6]_{J/\Psi} + N$ reaction. Unlike the predictions for the photo-production of $[^3He]_{J/\Psi}$ described in the previous subsection, very
little information on $[q^6]$ and the $ [q^6]-{J/\Psi}$ interaction
is available. We thus need to make various assumptions which can only be considered to be plausible for estimating the production cross sections.

In the impulse approximation, as described in section II, we need  the initial $N$-$[q^6]$ wavefunction in $^3He$ and the final $J/\Psi$-$[q^6]$
wavefunction to calculate the cross section of $\gamma + ^3He \rightarrow [q^6]_{J/\Psi} + N$.
In the following subsections, we explain our procedure for modeling these two ingredients of our predictions.

\subsubsection{Wavefunction of $N$-$q^6$ in $^3He$}
We start with a formulation of Refs.\cite{fasano-1,lee-matsuyama} within which the
 Hamiltonian for a two nucleon system is written as
\begin{eqnarray}
H= H_0 + v_{NN} + \sum_{\alpha}h_{[q^6]^{\alpha}\leftrightarrow NN}\,,
\label{eq:h-cbm}
\end{eqnarray}
where $\alpha$ denotes collectively the total angular momentum $J$, the total isospin $T$, and the parity $P$, and $v_{NN}$ is
a meson-exchange nucleon-nucleon ($NN$)
interaction. The vertex interaction $h_{[q^6]^{\alpha}\leftrightarrow NN}$ defines
the formation of a six-quark state $[q^6]^\alpha$ in $NN$ collisions.
The six-quark states $[q^6]^{\alpha}$ are identified with
the states predicted by the Bag model calculations of
 Mulder\cite{mulder}. By appropriately choosing the form of the vertex interaction $h_{[q^6]^{\alpha}\leftrightarrow NN}$, the NN scattering amplitudes derived from the Hamiltonian Eq.(\ref{eq:h-cbm}) are identical to those given by using the P-matrix approach of Jaffe and Low\cite{low}
and the Compound Bag Model formulation developed in Refs.\cite{itep,bakker}.

We will make use of the results of Fasano and Lee\cite{fasano-1,fasano-2}.
They determined the
mass $M_{\alpha}$ of $[q^6]^\alpha$ cluster and the interaction
$h_{[q^6]^{\alpha}\leftrightarrow NN}$ for $\alpha = ^1S_0$ and
$^3S_1$ by fitting the $NN$ scattering phase shifts up to 1 GeV. Within the simple s-wave
harmonic oscillator model for $^3He$, the probabilities
$P_{[q^6]^\alpha}$ of finding the $[q^6]^\alpha$-$N$ in $^3He$ are
estimated\cite{fasano-2} to be $P_{[q^6]^{^1S_0}}=0.7 \%$ and $P_{[q^6]^{^3S_1}}=0.06 \%$. For simplicity, we neglect the small
$^3S_1$ component. The bare mass of $[q^6]$ determined in Ref.\cite{fasano-1}
is $M_{^1S_0}=2150$ MeV. Here we will use these information
to  model the relative wavefunction of $[q^6]^\alpha$-$N$ which is
 needed to calculate the cross sections of $\gamma +^3He \rightarrow [q^6]_{J/\Psi} + N$.

\begin{figure}[htbp] \vspace{-0.cm}
\begin{center}
\includegraphics[width=0.6\columnwidth]{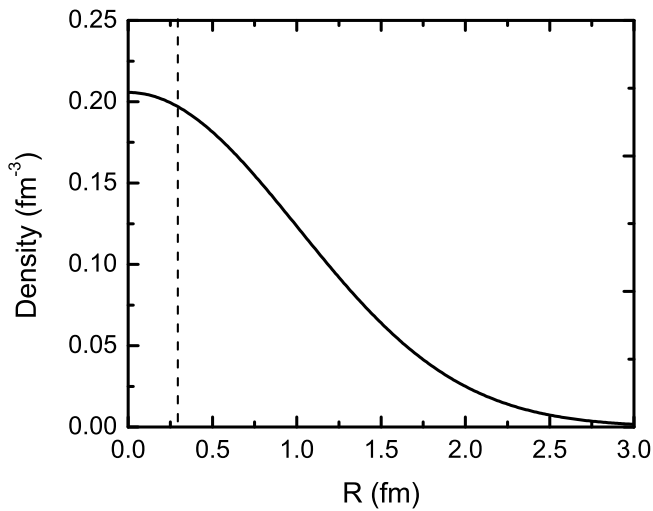}
\caption{The solid curve is the  normalized density distribution
calculated from using Eqs.(\ref{eq:rho1})-(\ref{eq:rho12})
with $b=1.35$ fm and $l_c=0.5$ fm.
  The dashed line defines  $r_c = 0.29$ fm for obtaining
the probability
$P_{[q^6]^{^1S_0}}=0.7\%$ for
finding the $[q^6]-N$ component in $^3He$ by using Eq.(\ref{eq:prob-6q}).}
\label{fg:denstyrange}
\end{center}
\end{figure}

\begin{figure}[htbp] \vspace{-0.cm}
\begin{center}
\includegraphics[width=0.49\columnwidth]{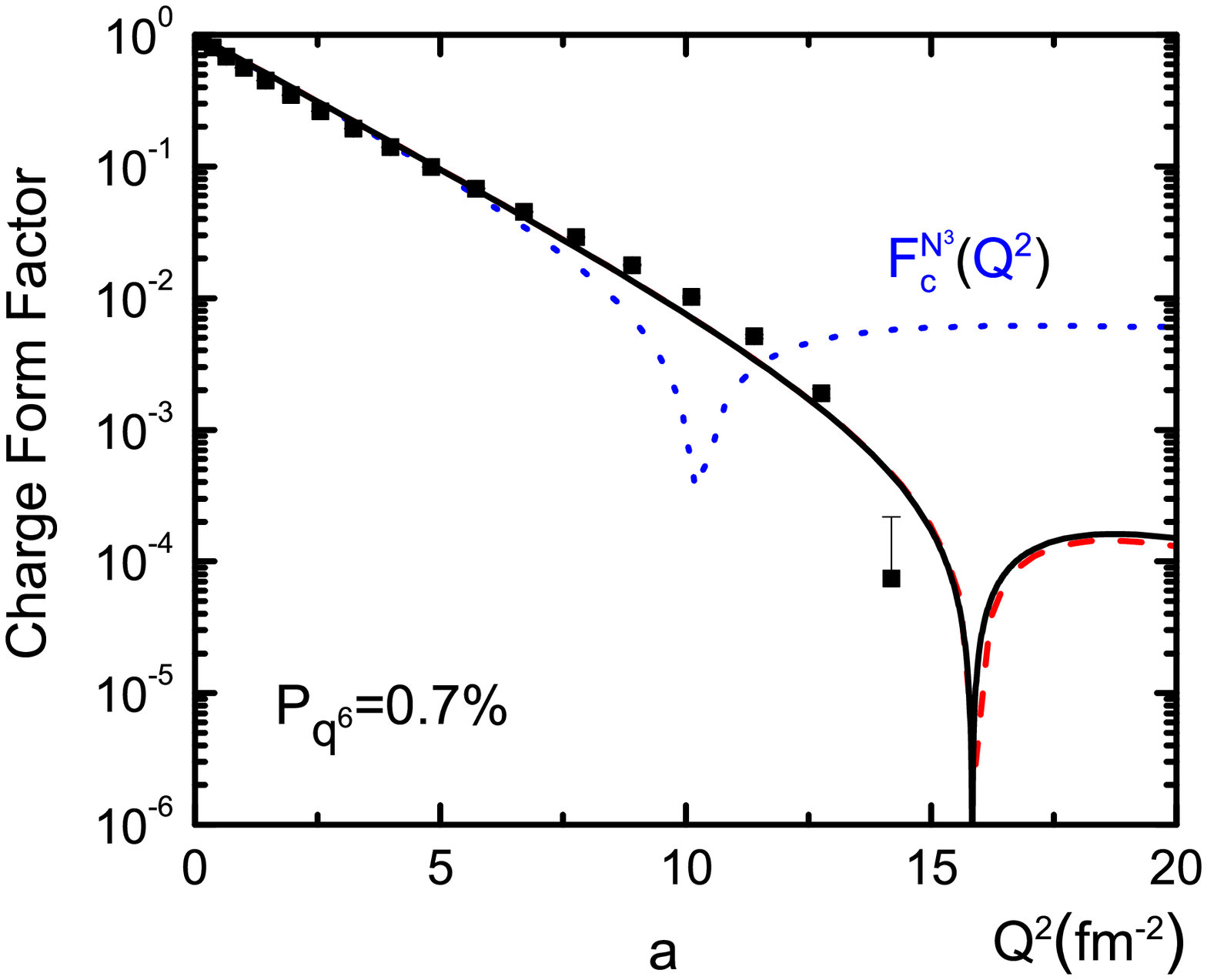}
\includegraphics[width=0.49\columnwidth]{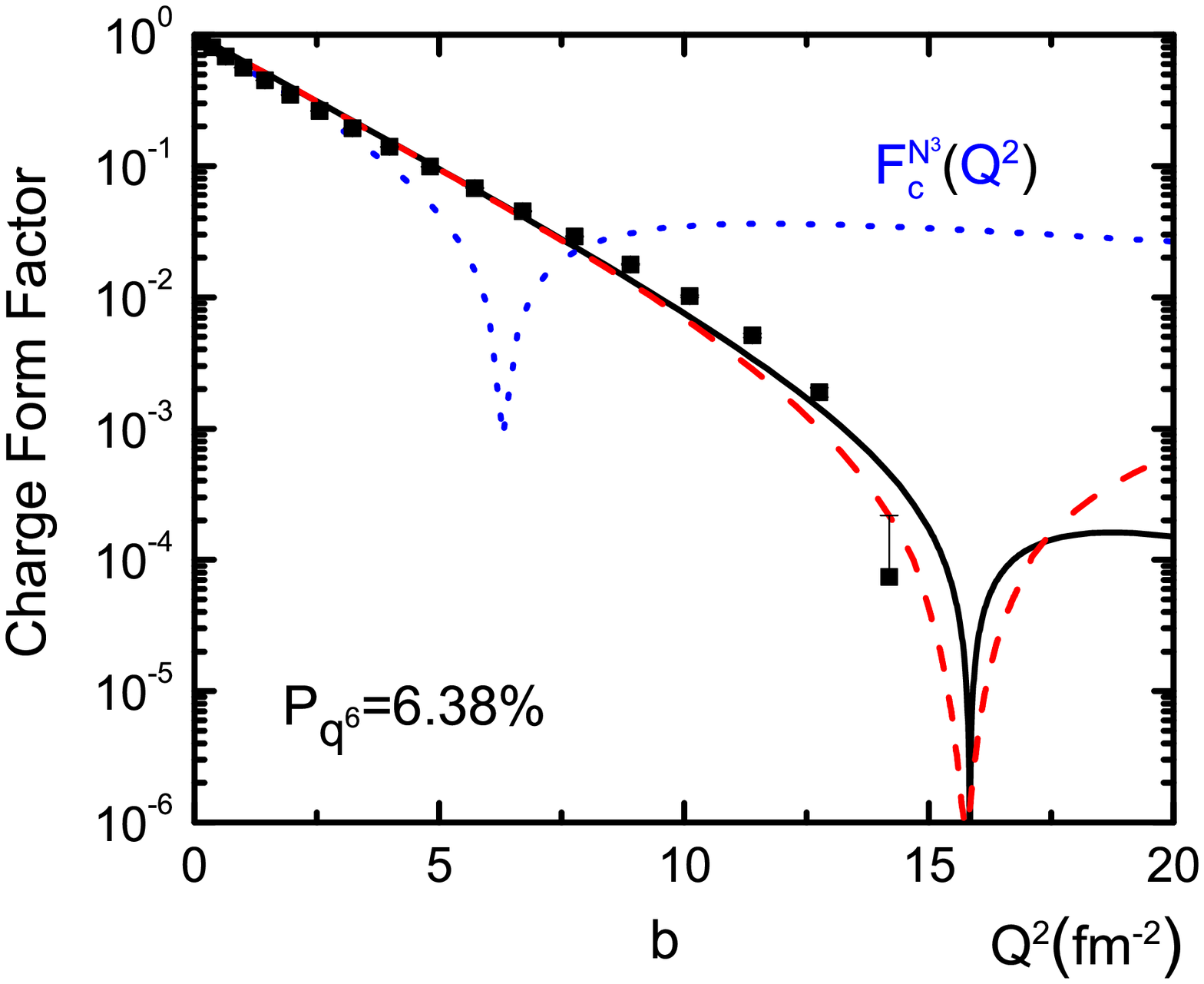}
\caption{The charge form factor $F_c(Q^2)$ for $^3He$.
The solid squares are from the Impulse Approximation (IA) calculation
of Ref.\cite{schiavilla}. The dotted blue curves
are $F^{N^3}_c(Q^2)$ calculated from using Eq.(\ref{eq:fcq-cut}) with
$r_c= 0.292$ fm and $P_{q^6} = 0.7\%$ (left) and $r_c=0.630$ fm and
$P_{q^6} = 6.3\%$ (right).
The red dashed curves are from adding the $q^6-N$ contributions
calculated from using Eq.(\ref{eq:form6q}) with $b^* = 0.185$ fm (left) and
$b^*=0.414$ fm (right).}
 \label{fig:formfactor}
\end{center}
\end{figure}

We assume that the charge distribution in the region with the distance $r\leq r_c$ from the center of $^3He$
is completely due to $[q^6]^\alpha$-$N$ components of the  wavefunction.
This is illustrated in Fig.\ref{fg:denstyrange}. Each $r_c$ clearly corresponds to a choice of $P_{q^6}$.
Within such a model, the charge form factor of $^3He$ is written as
\begin{eqnarray}
F_c(Q^2) = F_c^{N^3}(Q^2)+F_c^{q^6-N}(Q^2).
\label{eq:fcq-tot}
\end{eqnarray}
We next observe that within the conventional nuclear
model\cite{schiavilla}, the impulse approximation (IA) calculation,
which includes only the one-body nucleon current,
of $F_c(Q^2)$ is very close to the
data in $Q^2 \leq $ about 10 fm$^{-2}$ and can be reproduced very well by the Gaussian distribution
of the s-wave harmonic oscillator wavefunction.
  The IA results from
Ref.\cite{schiavilla} in this $Q^2$ region are the solid squares in Fig.\ref{fig:formfactor}.
We next demand  that the s-wave three-nucleon wavefunction reproduce these IA results.
 In addition, the resulting $F_c(Q^2)$ in the
higher $Q^2$ region
must have the similar structure of IA up $Q^2 \sim 20$ fm$^{-2}$ although we
do not have
higher partial wave components of the three-nucleon wavefunction.
We achieve this by using the s-wave harmonic oscillator wavefunction with Jastrow two-body correlation
used in Refs.\cite{feshbach,lee}. We write
\begin{eqnarray}
F_c(Q^2)&=&\int e^{-i\vec{Q}\cdot\vec{r}}\rho(\vec{r}) d\vec{r}
\end{eqnarray}
with
\begin{eqnarray}
\rho(\vec{r}_1)=\int d\vec{r}_2 \rho_2(\vec{r}_1,\vec{r}_2) \,, \label{eq:rho1}
\end{eqnarray}
where the two-body density is defined by
\begin{eqnarray}
\rho_2(\vec{r}_1, \vec{r}_2)&=&Ne^{-\frac{r^2_1+r^2_2}{2b^2}}(1-e^{-\frac{|\vec{r}_1-\vec{r}_2|^2}{2l_c^2}})\,,\nonumber\\
N&=&\frac{1}{(\sqrt{\pi}b)^3}\sqrt{\frac{[l_c^2+b^2]^{3/2}}{[l_c^2+b^2]^{3/2}-l_c^3}}\,. \label{eq:rho12}
\end{eqnarray}
As seen in Fig.\ref{fig:formfactor}, the solid black curve calculated with $b=1.35$ fm and $l_c=0.5$ fm
can reproduce the impulse approximation calculation results (solid squares) given
in Ref.\cite{schiavilla} up to  $Q^2 \sim 10$ fm$^{-2}$. At higher $Q^2$, the solid curves
have the similar structure of the IA results. For our present s-wave calculations, we consider
the solid curves in Fig.\ref{fig:formfactor} as the $F_c(Q^2)$ in Eq.(\ref{eq:fcq-tot}).
Accordingly,
the $NNN$ contribution in Eq.(\ref{eq:fcq-tot}) is calculated from
\begin{eqnarray}
F_c^{N^3}(Q^2) = \int^\infty_{r_{c}}r^2 dr\int d\Omega_{r}
e^{-i\vec{Q}\cdot\vec{r}} \rho(\vec{r}) \,
\label{eq:fcq-cut}
\end{eqnarray}
and
 the probability $P_{[q^6]}$ is defined by
\begin{eqnarray}
P_{q^6}&=&\int^{r_{c}}_0 r^2 d{r}\int d\Omega_{r}\rho(\vec{r})
\label{eq:prob-6q} \,.
\end{eqnarray}

For $P_{q^6}=0.7\%$ determined in Ref.\cite{fasano-1} within the Compound Bag Model of NN scattering,
we choose $r_c=0.292$ fm to calculate Eq.(\ref{eq:fcq-cut}) and
 get the blue dotted curve in the left-hand side of Fig.\ref{fig:formfactor}.
In the right-hand side, the blue dotted curve is from the calculation using
Eq.(\ref{eq:fcq-cut}) with $r_c=0.630$ fm which gives
$P_{q^6}=6.3\%$.
Clearly, both results agrees well with the IA (solid squares) and the solid curve
 only in the  low $Q^2 $ region.
Our next task is to model $F^{q^6-N}(Q^2)$ such that for each $r_c$, $F_c(Q^2)$ (solid black curve) in  Fig.\ref{fig:formfactor} up to $Q^2 \sim 15$ fm$^2$ can be reproduced, as required by Eq.(\ref{eq:fcq-tot}).

For simplicity, we assume that $F_c^{q^6-N}(Q^2)$ can  be calculated from a normalized Gaussian distribution
\begin{eqnarray}
F_c^{q^6-N}(Q)&=P_{q^6}&\int e^{-i\vec{Q}\cdot\vec{r}}
[\frac{1}{(\sqrt{\pi}b^*)^{3} }e^{\frac{-r^2}{b^{*2}}}]d \vec{r}\,.
\label{eq:form6q}
\end{eqnarray}
Accordingly, the mean radius of $q^6$-$N$ can be defined by
\begin{eqnarray}
<r^2> = \int
[\frac{1}{(\sqrt{\pi}b^*)^{3} }e^{\frac{-r^2}{b^{*2}}}]r^2 d \vec{r}\,.
\label{eq:mrdii}
\end{eqnarray}

We adjust $b^*$ for each $P_{q^6}$ to fit the solid curves
 in Fig.\ref{fig:formfactor}.
We find that if $P_{q^6}$ is larger than $6.5\%$,
 no $b^*$ can fit the form factor defined by Eq.(\ref{eq:fcq-tot}) in the $Q^2 < 20 $ fm$^{-2}$ region.
In the cluster model of Ref.\cite{japan-he3}, $P_{q^6} =4.0\%$ is obtained from fitting the
$^3He$ form factor. The $P_{q^6} = 15\%$ determined in Ref.\cite{vary} by fitting the structure function of $^3He(e,e')$ is beyond what our formulation can accommodate. For comparison, we thus choose three different models with
$P_{q^6}= 0.7\%, 4\%$, and $6.38\%$ for our calculations. In Table \ref{all},
we list $r_c$, $b^*$, and also $<r^2>^{1/2}$ calculated from using
Eq.(\ref{eq:mrdii}) for these three cases. Our fits
are the red dashed curves in Fig.\ref{fig:formfactor}
for $P_{q^6}= 0.7\%$ (left)
and $6.38\%$ (right).

Once $b^*$ is determined, we then
 assume that the relative wavefunction of $[q^6]$-N  can be described by the harmonic wavefunction with the same $b^*$. This should be
reasonable for making order of magnitude estimates in this work. A more sophisticated approach should
account for the quark charge distribution in $[q^6]$ which is beyond the scope of this work.
Also, the sharp cutoff at $r=r_c$ to define $F_c^{N^3}$ in Eq.(\ref{eq:fcq-cut})
 should perhaps be
better modeled. For our present qualitative estimations, this simple procedure should be sufficient.

\begin{table}[ht]
\begin{center}
\caption{The parameters for $^6q$-$N$ and $^6q$-$J/\psi$ systems.
See text for the explanations of the notations.}\label{all}
\begin{tabular}{ccccc|ccc}\hline
     &\multicolumn{3}{c}{$q^6$-$N$} &      &  \multicolumn{3}{c}{$q^6$-$J/\psi$} \\ \hline
 Model &  $P_{q^6}$     &  $r_c$  & $b^*$  & ${<r^2>^{1/2}}$  & $\mu_{q^6}$   & $\alpha_{q^6}$ & B.E. \\
       &                &    (fm) &   (fm) &   (fm)           &   (GeV)       &                & (MeV)\\
$A_1$ & 0.7\%  &  0.292 & 0.185 & 0.226          & 0.6        & 1.33 & 498.42\\
$A_2$ &        &        &       &                & 1.0    & 1.50 & 389.63\\
&&&&&&&\\
$B_1$ & 4.0\%  &  0.533 & 0.346 & 0.424          & 0.6     & 0.83  & 104.08\\
$B_2$ &        &        &       &                & 1.0    & 1.05& 79.77\\
&&&&&&&\\
$C_1$ & 6.38\% &  0.630 & 0.414 & 0.507          & 0.6     & 0.75  & 65.84\\
$C_2$ &        &        &       &                & 1.0    & 0.97  & 51.33\\
\hline
\end{tabular}
\end{center}
\end{table}

\subsubsection{Wavefunction of $^6q$-$J/\psi$ bound state}

We follow the procedure of subsection IV.B to assume that the
$q^6$-$J/\Psi$ bound states ($[q^6]_{J/\Psi}$)
 are also defined by a potential,
of Yukawa form
\begin{eqnarray}
V_{J/\Psi,q^6}(r)&=&- \alpha_{q^6}\frac{e^{-\mu_{q^6}r}}{r}\label{yukawa}\,.
\label{eq:yukawa-jpsi}
\end{eqnarray}
We expect that if a $[q^6]_{J/\Psi}$ bound state can be
produced, its size must be small  for
color field to give strong attractive force. Thus it is reasonable to
assume that the mean radius of $[q^6]_{J/\Psi}$
is close to the value $<r^2>^{1/2}$ of the initial $q^6$-$N$ system listed in Table \ref{all}.
We find that such a small size can be generated from
choosing $\mu_{q^6} > 0.6 $ GeV in defining the potential
Eq.(\ref{eq:yukawa-jpsi}). Once a value of $\mu_{q^6} $ is chosen,
we then determine the potential strength $\alpha_{q^6}$ by requiring
\begin{eqnarray}
<r^2> = \int |\phi_{q^6,J/\Psi}(\vec{r})|^2 r^2 d\vec{r}\,,
\end{eqnarray}
where $\phi_{q^6,J/\Psi}(\vec{r})$ is the $J/\Psi$-$q^6$
relative wavefunction generated from the potential Eq.(\ref{eq:yukawa-jpsi}),
and the values of $<r^2>^{1/2}$ for various considered cases
are listed in Table \ref{all}. The resulting $\alpha_{q^6}$  and the binding energies (B.E.) are also listed there.
Here we note that the binding energy increases as the mean radius
$<r^2>^{1/2}$
and the corresponding probability
$P_{q^6}$ decrease.

\subsubsection{The Results of $\gamma + ^3He \to [q^6]_{J/\psi} + N$}

With the wavefunctions for $q^6$-$N$  and $q^6$-$J/\Psi$ specified in the previous subsections, we can use the formula in section III, with trivial changes in
notations and spin quantum numbers,
 to calculate the total cross section of
$\gamma + ^3He \to [q^6]_{J/\psi} + N$. However, we need to
 multiply the results by the probability $P_{q^6}$ of the $N$-$[q^6]$
component in $^3He$; namely the results from using Eq.(\ref{eq:d-prod-1})
is changed to
\begin{eqnarray}
\frac{d\sigma}{d\Omega} \rightarrow P_[q^6] \times [\frac{d\sigma}{d\Omega}]_0\,,
\label{eq:crst-q6}
\end{eqnarray}
where $[\frac{d\sigma}{d\Omega}]_0$ is calculated from using
Eq.(\ref{eq:d-prod-1}) and all subsequent equations in section III.A.

\begin{figure}[htbp] \vspace{-0.cm}
\begin{center}
\includegraphics[width=0.49\columnwidth]{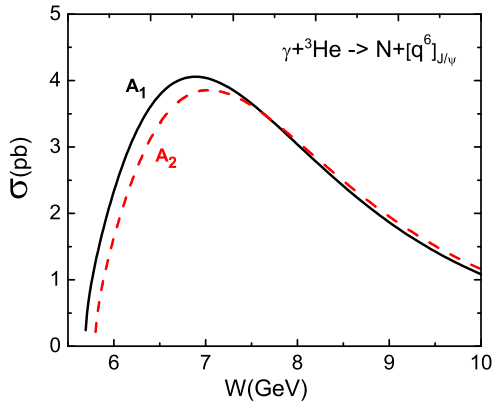}
\caption{
The total cross section of $\gamma + ^3He \to [q^6]_{J/\psi} + N$
as function of the $\gamma$-$^3He$ invariant mass $W$.
The black solid and red dashed curves are for the case $A_1$ and $A_2$ in the
Table \ref{all}, respectively.
} \label{fg:6q-A1A2}
\end{center}
\end{figure}

 We first consider the case that $P_{[q^6]}=0.7\%$ as determined in
Refs.\cite{fasano-1,fasano-2} from fitting the $NN$ phase shifts up to 1 GeV.
By using the parameters for models $A_1$ and $A_2$ listed in Table \ref{all},
we obtain the results shown in
Fig.\ref{fg:6q-A1A2}.
We observe that with the same small radius $<r^2>^{1/2}=0.226$ fm for the produced
$[q^6]_{J/\psi}$ system, the predicted cross
sections are very close despite their potential range, measured by
 $1/\mu_{q^6}$, and coupling constant $\alpha_{q^6}$ can be very different.
The same finding is also from comparing the predicted cross sections
from the models $B_1$ and $B_2$, and also the models $C_1$ and $C_2$.

\begin{figure}[htbp] \vspace{-0.cm}
\begin{center}
\includegraphics[width=0.49\columnwidth]{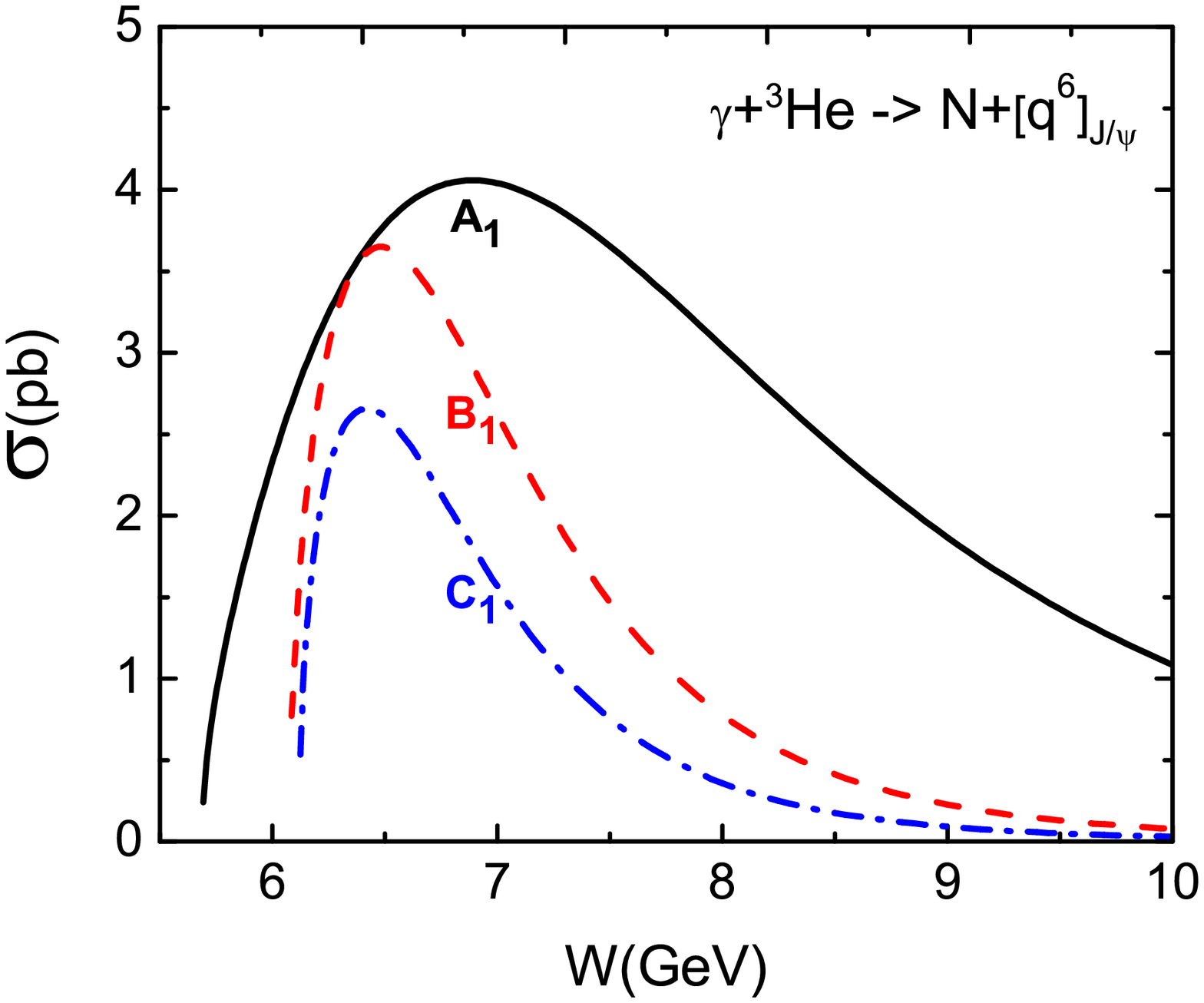}
\includegraphics[width=0.49\columnwidth]{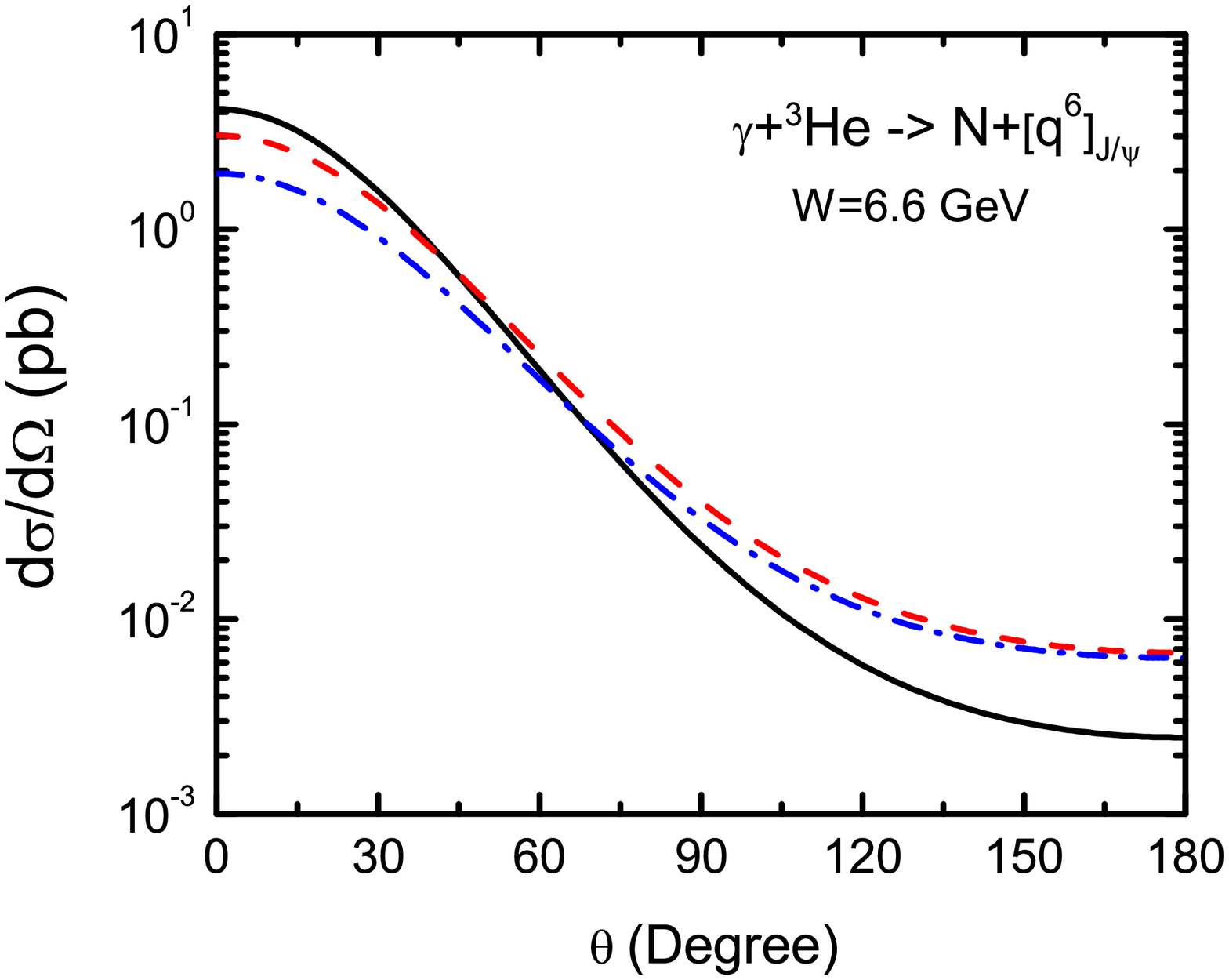}
\caption{The black solid, red dashed and blue dotted-dashed lines are
from Models $A_1$, $B_1$, and $C_1$ of Table \ref{all},
respectively.
Left:
The total cross section of $\gamma + ^3He \to [q^6]_{J/\psi} + N$
as function of the $\gamma$-$^3He$ invariant mass $W$.
Right:The differential cross section of $\gamma + ^3He \to q^6_{[J/\psi]} + N$  of
the outgoing $N$ at  $\gamma$-$^3He$
invariant mass $W= 6.6$ GeV. }
 \label{fg:diffABC}
\end{center}
\end{figure}

In the left-hand side of
Fig.\ref{fg:diffABC}, we show the dependence
 of the predicted cross sections on $P_{q^6}$ by comparing the cross
sections from three models $A_1$, $B_1$, and
$C_1$ listed in
Table \ref{all}. We observe that as $P_{q^6} $ decreases, the peak is shifted to higher energies. Each case has different threshold energy
 due to their differences in binding energies, as seen in Table \ref{all}. Their magnitudes are comparable despite their $P_{q^6}$ are very different.
 We find that this is due to the fact that the cross section
$[\frac{d\sigma}{d\Omega}]_0$ in Eq.(\ref{eq:crst-q6}) for the
model with smaller $P_{q^6}=0.7\%$  is a factor  of about 10 larger than that
for the model with larger $P_{q^6}= 6.38 \%$, since  this large momentum
transfer reaction favors the production of $[q^6]_{J/\Psi}$ with smaller
size characterized by $<r^2>^{1/2}$ in Table \ref{all}.
 The situation is
similar to what we discussed in explaining the results shown in Fig.\ref{fig:compare}.
Thus the  magnitudes of the cross sections from three models at peak
positions are comparable because the factor of about 10 difference
in $[\frac{d\sigma}{d\Omega}]_0$ in Eq.(\ref{eq:crst-q6}) is compensated by the
similar factor of about 10 in  $P_{q^6} $.
However,
 the three models have rather different energy dependence, as also seen in
the left-hand side of Fig.\ref{fg:diffABC}.
On the other hand, they are all forward peaked, as shown in the right-hand side of
 Fig.\ref{fg:diffABC} for the differential cross sections at $W=6.6$ GeV.

The results shown in Fig.\ref{fg:diffABC} suggest that the upper bound of
 the predicted total cross sections of
$\gamma + ^3He \rightarrow [q^6]_{J/\Psi} +N$ is about 2 - 4 pico-barn

\section{summary and discussions}
We have presented predictions of the cross sections of
$\gamma+ ^4He \rightarrow N + [^3He]_{J/\Psi}$ reaction at energies near the
 $J/\Psi$ production threshold.
In the impulse approximation, the calculations have been performed by
using several $\gamma+N \rightarrow J/\Psi +N$ models based on the
Pomeron-exchange and pion-exchange mechanisms.
The $J/\Psi$ wavefunctions in $[^3He]_{J/\Psi}$ are generated from
various $J/\Psi$-nucleus potentials which are
constructed by  either using a procedure based on the
Pomeron-quark coupling mechanism\cite{brodsky90},
or folding a $J/\Psi$-N potential $v_{J/\Psi,N}$ into the nuclear densities.
We consider $v_{J/\Psi,N}$ derived from the effective field theory approach, Lattice QCD,
and Pomeron-quark coupling model. The upper bound of the predicted total cross
sections is about 0.1 - 0.3 pico-barn.

Clearly, our investigations are only
for estimating the cross sections to facilitate the
experimental considerations of
 possible measurements of $[^3He]_{J/\Psi}$ bound states
 at Jefferson Laboratory.
Several improvements are needed for more quantitative predictions.
First we need precise data of $\gamma+p \rightarrow J/\Psi +p$ near threshold to
distinguish several models we have considered and also to develop a more
sophisticated model.We also need the data to pin down
 the $J/\Psi$-N interaction for a more realistic
calculation of $J/\Psi$-nucleus potential such as the folding model considered in this work.
One possibility is to use the $\pi +^2H \rightarrow p + J/\Psi + n$
reaction to extract the $J/\Psi$-$N$ scattering length, as suggested in Ref.\cite{brodsky-1}.
Alternatively, we can apply the model presented in this paper to
 determine the $J/\Psi$-$N$ interactions
by investigating
 the $\gamma +^2H \rightarrow p + J/\Psi + n$ reaction. Possible experiments on these two
processes will be very useful.
We of course also need to
use more realistic wavefunctions for $^3He$ and $^4He$ while the s-wave oscillator
wavefunctions employed in this investigation are reasonably consistent
with the charge form factors calculated from
the conventional nuclear models.

Motivated by the previous
investigations\cite{brodsky-3,brodsky-2} on the effects due to
multi-quark clusters in $pp$ and $\gamma+^2H\rightarrow J/\Psi +n + p $,
we have also considered the possibility of the production
of a  $[q^6]-J/\Psi$ bound state due to a six-quark $[q^6]$ cluster in
$^3He$. The Compound Bag Model of $NN$ scattering and the quark cluster model of nuclei are
used to estimate  the $[q^6]$-N  wavefunction in $^3He$ by imposing the condition that
the sum of the contributions from $[q^6]$-N and $NNN$ components to the
$^3He$ charge form factor
must be consistent with what are predicted by the conventional
nuclear models\cite{schiavilla}  which explain the data very well.
The upper bound of the predicted total cross sections of
$\gamma + ^3He \rightarrow [q^6]_{J/\Psi} +N$ is about 2 - 4 pico-barn, depending on the
model of $\gamma+N \rightarrow J/\Psi +N$ used in the calculations.
If such bound states can be identified, it will open up a new window for investigating the
role of the gluon field in determining the hadron structure.

\begin{acknowledgments}
We thank Kawtar Hafidi for the discussions on the possible $J/\Psi$ production
experiments
at Jeferson Laboratory, Henning Esbensen and Rocco Schiavilla for their help
in our bound state calculations.
This work is supported by the U.S. Department of Energy, Office of Nuclear Physics Division,
under Contract No. DE-AC02-06CH11357.
This research used resources of the National Energy Research Scientific Computing Center,
which is supported by the Office of Science of the U.S. Department of Energy
under Contract No. DE-AC02-05CH11231, and resources provided on ``Fusion,''
a 320-node computing cluster operated by the Laboratory Computing Resource Center
at Argonne National Laboratory.
\end{acknowledgments}


\begin{thebibliography}{10}
\bibitem{pesk79}
M.E. Peskin, Nucl. Phys. {\bf B156}, 365 (1979)

\bibitem{bp79}
G. Bhanot and M.E. Peskin,  Nucl. Phys. {\bf B156}, 391 (1979)

\bibitem{luke92}
M. Luke, A.V. Manohar, and M.J. Savage Phys. Lett B {\bf 288}, 355 (1992)

\bibitem{brodsky-1}
  S.~J.~Brodsky and G.~A.~Miller,
   Phys.\ Lett.\ B {\bf 412}, 125 (1997).

\bibitem{lqcd}
Taichi Kawanai and Shoichi Sasaki, Phys. Rev. D {\bf 82}, 091501 (2010)

\bibitem{russia}
A.B. Kaidalov and P.E. Volkovitsky, Phys. Rev. Lett {\bf 69}, 3155 (1992)


\bibitem{brodsky90}
S.J. Brodsky, I.A. Schmidt, and G.F. Teramond, Phys. Rev. Lett.
{\bf 64}, 1011 (1990)

\bibitem{donn84}
A. Donnachie and P.V. Landshoff, Nucl. Phys. {\bf B244}, 322 (1984)

\bibitem{feshbach-1}
Herman Feshbach, {\bf Theoretical Nuclear Physics, Nuclear Reactions}
(Wiley, New York, 1992)


\bibitem{workshop}
 Z.-E. Meziani, K. Hafidi, X. Uian, and N. Sparveris et al.,
Proposal "Near Threshold Electroproduction of $J/\Psi$ at 11 GeV",
PR12-12-006(2012), PAC39, Jefferson Laboratory (2012)



\bibitem{brodsky-3}
S.~J.~Brodsky and G.F. de Teramond, Phys. Rev. Lett {\bf 60}, 1924 (1988)

\bibitem{brodsky-2}
  S.~J.~Brodsky, E. Chudakov, P. Hoyer, and J.M. Laget, Phys. Lett
{\bf B498}, 23 (2001)


\bibitem{low}
R.L. Jaffe and F. Low, Phys. Rev. D {\bf 19}, 2105 (1979); F. Low in
{\bf Pointlike Structure Inside and Outside Hadrons}, Proceedings of
1979 Erice Summer School, Editted by A. ZiChiChi (Plenum, New York, 1979),
p. 155.

\bibitem{mulder}
P. J. Mulders. Phys. Rev. D {\bf 26}, 3039 (1982); {\bf 28}, 443 (1983)

\bibitem{itep}
Yu. A. Simonov, Phys. Lett. {\bf 107B}, 1 (1981); Yad. Fiz {\bf 38},
1542 (1983)[Sov. J. Nucl. Phys. {\bf 38}, 939 (1983)

\bibitem{bakker}

B.L.G. Bakker, I.L. Grach, and I.M. Narodetskii, Nucl. Phys. {\bf A424},
563 (1984)

\bibitem{fasano-1}
C. Fasano and T.-S. H. Lee, Phys. Rev. C {\bf 36}, 1906 (1987)

\bibitem{fasano-2}
C. Fasano and T.-S. H. Lee, Phys. Lett. {\bf 271B}, 9 (1989)


\bibitem{vary}
H.J. Pirner and J.P. Vary, Phys. Rev. Lett. {\bf 46},1376 (1981)


\bibitem{japan-he3}
M. Namiki, K. Okano, and N. Oshimo, Phys. Rev. C {\bf 25}, 2157 (1982).

\bibitem{LN87}
P.~V. Landshoff and O.~Nachtmann,
  Z. Phys. C {\bf 35}, 405 (1987).

\bibitem{LM95}
J.-M. Laget and R.~Mendez-Galain,
  Nucl. Phys. {\bf A581}, 397 (1995).

\bibitem{PL97}
M.~A. Pichowsky and T.-S.~H. Lee,
  Phys. Rev. D {\bf 56}, 1644 (1997).

\bibitem{titov-lee}
A. I. Titov, T.-S. H. Lee, Phys. Rev. C {\bf 67}, 065205 (2003).

\bibitem{ky12}
Alvin Kiswandhi and Shin Nan Yang,
Phys.Rev. C {\bf 86}, 015203 (2012), Erratum-ibid. C {\bf 86}, 019904 (2012)




\bibitem{pdg}
J. Beringer  et al.
 (Particle Data Group), Phys. Rev. D {\bf 86}, 010001 (2012)

\bibitem{OL02}
Y.~Oh and T.-S.~H. Lee,
  Phys. Rev. C {\bf 66}, 045201 (2002).

\bibitem{gao01}
H. Gao, T.-S. H. Lee, and V. Marinov, Phys. Rev. C {\bf 63}, 022201 (2001)






\bibitem{Binkley:1981kv}
  M.~E.~Binkley, C.~Bohler, J.~Butler, J.~P.~Cumalat, I.~Gaines, M.~Gormley, D.~Harding and R.~L.~Loveless {\it et al.},
  Phys.\ Rev.\ Lett.\  {\bf 48}, 73 (1982).  

\bibitem{Denby:1983az}
  B.~H.~Denby, V.~K.~Bharadwaj, D.~J.~Summers, A.~M.~Eisner, R.~G.~Kennett, A.~Lu, R.~J.~Morrison and M.~S.~Witherell {\it et al.},
  Phys.\ Rev.\ Lett.\  {\bf 52}, 795 (1984).  

\bibitem{Barate:1986fq}
  R.~Barate {\it et al.}  [NA14 Collaboration],
  Z.\ Phys.\ C {\bf 33}, 505 (1987).  

\bibitem{Frabetti:1993ux}
  P.~L.~Frabetti {\it et al.}  [E687 Collaboration],
  Phys.\ Lett.\ B {\bf 316}, 197 (1993).  

\bibitem{Camerini:1975cy}
  U.~Camerini, J.~G.~Learned, R.~Prepost, C.~M.~Spencer, D.~E.~Wiser, W.~Ash, R.~L.~Anderson and D.~Ritson {\it et al.},
  Phys.\ Rev.\ Lett.\  {\bf 35}, 483 (1975).  

\bibitem{Gittelman:1975ix}
  B.~Gittelman, K.~M.~Hanson, D.~Larson, E.~Loh, A.~Silverman and G.~Theodosiou,
  Phys.\ Rev.\ Lett.\  {\bf 35}, 1616 (1975).  

\bibitem{Anderson}
  R. L. Anderson, Excess Muons and New Results in $\Psi$ Photoproduction. SLAC-PUB-1471 (unpublished).

\bibitem{H1}
  S.~Aid {\it et al.}  [H1 Collaboration],
  Nucl.\ Phys.\ B {\bf 472}, 3 (1996).
  A.~Aktas {\it et al.}  [H1 Collaboration],
  Eur.\ Phys.\ J.\ C {\bf 46}, 585 (2006).

\bibitem{ZEUS}
  J.~Breitweg {\it et al.}  [ZEUS Collaboration],
  Z.\ Phys.\ C {\bf 76}, 599 (1997).
  S.~Chekanov {\it et al.}  [ZEUS Collaboration],
  Eur.\ Phys.\ J.\ C {\bf 24}, 345 (2002).
  S.~Chekanov {\it et al.}  [ZEUS Collaboration],
  Nucl.\ Phys.\ B {\bf 695}, 3 (2004).
  M.~Derrick {\it et al.}  [ZEUS Collaboration],
  Phys.\ Lett.\ B {\bf 350}, 120 (1995).



\bibitem{nucl-data}
R. Hofstadter, Annu. Nucl. Sci. {\bf 7}, 231 (1957);
J.S. McCarthy et al., Phys. Rev. C {\bf 15}, 1396 (1977).



\bibitem{bohr-moltt}
Aage Bohr and Ben R. Mottelson, {\bf Nuclear Structure} Volume I, 1969
(W.A. Benjamin, Inc)

\bibitem{schiavilla}
J.\ Carlson and R.\ Schiavilla,
Rev.\ Mod.\ Phys.\ {\bf 70}, 743 (1997);L.E.\ Marcucci, D.O.\ Riska, and R.\ Schiavilla
Phys.\ Rev.\ C {\bf 58}, 3069 (1998).


\bibitem{lee-matsuyama}
T.-S. H. Lee and A. Matsuyama, Phys. Rev. C {\bf 32}, 516 (1985)


\bibitem{feshbach}
H. Feshbach, A. Gal, and J. Hufner, Ann. Phys. (N.Y.) {\bf 66}, 20 (1971)

\bibitem{lee}
T.-S. H. Lee and S. Chakravarti,
Phys. Rev. C {\bf 16}, 273 (1977)

\end{thebibliography}
\end{document}